\documentclass[twocolumn]{aastex631}
\pdfoutput=1

\usepackage{appendix}
\usepackage{overpic}

\newcommand{\Msun}{M\textsubscript{\(\odot\)}}
\newcommand{\citediamondback}{Morley et al., 2024,in preparation}

\submitjournal{\aj}
\accepted{March 4, 2024}
\published{April 9, 2024}

\shortauthors{Tobin et al.}

\begin{document}

\title{Direct-imaging Discovery of a Substellar Companion Orbiting the Accelerating Variable Star, HIP~39017}

\correspondingauthor{T. L. Tobin}
\email{tltobin@umich.edu}

\author[0000-0001-8103-5499]{Taylor L. Tobin}
\affiliation{Department of Astronomy, University of Michigan, 1085 S. University, Ann Arbor, MI 48109, USA}
\affiliation{Department of Physics and Astronomy, University of Notre Dame,
Nieuwland Science Hall, Notre Dame, IN 46556, USA}

\author[0000-0002-7405-3119]{Thayne Currie}
 \affiliation{Subaru Telescope, National Astronomical Observatory of Japan, 
650 North A`oh$\bar{o}$k$\bar{u}$ Place, Hilo, HI  96720, USA}
\affiliation{Department of Physics and Astronomy, University of Texas at San Antonio, San Antonio, TX 78249, USA}

\author[0000-0002-6845-9702]{Yiting Li}
\affiliation{Department of Astronomy, University of Michigan, 1085 S. University, Ann Arbor, MI 48109, USA}
\affiliation{Department of Physics, University of California, Santa Barbara, Santa Barbara, CA 93106-9530, USA}

\author[0000-0001-6305-7272]{Jeffrey Chilcote}
\affiliation{Department of Physics and Astronomy, University of Notre Dame,
Nieuwland Science Hall, Notre Dame, IN 46556, USA}

\author[0000-0003-2630-8073]{Timothy D. Brandt}
\affiliation{Department of Physics, University of California, Santa Barbara, Santa Barbara, CA 93106-9530, USA}

\author[0000-0002-9420-4455]{Brianna Lacy}
\affiliation{Department of Astronomy \& Astrophysics, University of California, Santa Cruz, Santa Cruz, CA 95064, USA}

\author[0000-0002-4677-9182]{Masayuki Kuzuhara}
\affiliation{Astrobiology Center, NINS, 2-21-1 Osawa, Mitaka, Tokyo 181-8588, Japan}
\affiliation{National Astronomical Observatory of Japan, 2-21-1 Osawa, Mitaka, Tokyo 181-8588, Japan}

\author[0000-0001-5763-378X]{Maria Vincent}
\affiliation{Institute for Astronomy, University of Hawai'i at M\={a}noa, 2680 Woodlawn Dr, Honolulu, HI 96822, USA}

\author[0000-0001-9441-7656]{Mona El Morsy}
\affiliation{Department of Physics and Astronomy, University of Texas at San Antonio, San Antonio, TX 78249, USA}

\author[0000-0003-4514-7906]{Vincent Deo}
 \affiliation{Subaru Telescope, National Astronomical Observatory of Japan, 
650 North A`oh$\bar{o}$k$\bar{u}$ Place, Hilo, HI  96720, USA}

\author[0000-0001-5058-695X]{Jonathan P. Williams}
\affiliation{Institute for Astronomy, University of Hawai'i at M\={a}noa, 2680 Woodlawn Dr, Honolulu, HI 96822, USA}

\author[0000-0002-1097-9908]{Olivier Guyon}
 \affiliation{Subaru Telescope, National Astronomical Observatory of Japan, 
650 North A`oh$\bar{o}$k$\bar{u}$ Place, Hilo, HI  96720, USA}

\author[0000-0002-3047-1845]{Julien Lozi}
 \affiliation{Subaru Telescope, National Astronomical Observatory of Japan, 
650 North A`oh$\bar{o}$k$\bar{u}$ Place, Hilo, HI  96720, USA}

\author[0000-0003-4018-2569]{Sebastien Vievard}
 \affiliation{Subaru Telescope, National Astronomical Observatory of Japan, 
650 North A`oh$\bar{o}$k$\bar{u}$ Place, Hilo, HI  96720, USA}

\author[0000-0002-9372-5056]{Nour Skaf}
\affiliation{Department of Astronomy \& Astrophysics, University of California, Santa Cruz, Santa Cruz, CA 95064, USA} 

\author[0000-0002-1094-852X]{Kyohoon Ahn}
 \affiliation{Subaru Telescope, National Astronomical Observatory of Japan, 
650 North A`oh$\bar{o}$k$\bar{u}$ Place, Hilo, HI  96720, USA}

\author[0000-0001-5978-3247]{Tyler Groff}
\affiliation{NASA Goddard Space Flight Center, Greenbelt, MD 20770, USA}
\author[0000-0002-6963-7486]{N. Jeremy Kasdin}
\affiliation{Department of Mechanical Engineering, Princeton University, Princeton, NJ 08544, USA}

\author[0000-0002-6879-3030]{Taichi Uyama}
\affiliation{Department of Physics and Astronomy, California State University, Northridge, Northridge, CA 91330, USA}
\author[0000-0002-6510-0681]{Motohide Tamura}
\affil{Astrobiology Center, NINS, 2-21-1 Osawa, Mitaka, Tokyo 181-8588, Japan}
\affiliation{National Astronomical Observatory of Japan, 2-21-1 Osawa, Mitaka, Tokyo 181-8588, Japan}
\affiliation{Department of Astronomy, Graduate School of Science, The University of Tokyo, 7-3-1, Hongo, Bunkyo-ku, Tokyo, 113-0033, Japan}

\author[0000-0002-9027-4456]{Aidan Gibbs}
\affil{Department of Physics \& Astronomy, University of California, Los Angeles, CA 90095, USA}

\author[0000-0002-8984-4319]{Briley L. Lewis}
\affil{Department of Physics \& Astronomy, University of California, Los Angeles, CA 90095, USA}

\author[0000-0001-5831-9530]{Rachel Bowens-Rubin}
\affiliation{Department of Physics, University of California, Santa Barbara, Santa Barbara, CA 93106-9530, USA}

\author[0000-0002-5082-6332]{Ma\"issa Salama}
\affiliation{Department of Physics, University of California, Santa Barbara, Santa Barbara, CA 93106-9530, USA}

\author[0000-0003-0115-547X]{Qier An}
\affiliation{Department of Physics, University of California, Santa Barbara, Santa Barbara, CA 93106-9530, USA}

\author[0000-0001-8892-4045]{Minghan Chen}
\affiliation{Department of Physics, University of California, Santa Barbara, Santa Barbara, CA 93106-9530, USA}

\begin{abstract}
We present the direct-imaging discovery of a substellar companion (a massive planet or low-mass brown dwarf) to the young, $\gamma$ Doradus ($\gamma$ Dor)-type variable star, HIP~39017 (HD~65526). The companion's SCExAO/CHARIS JHK ($1.1-2.4\mu$m) spectrum and Keck/NIRC2 L$^{\prime}$ photometry indicate that it is an L/T transition object. A comparison of the JHK$+$L$^{\prime}$ spectrum to several atmospheric model grids finds a significantly better fit to cloudy models than cloudless models. Orbit modeling with relative astrometry and precision stellar astrometry from Hipparcos and Gaia yields a semi-major axis of $23.8^{+8.7}_{-6.1}$~au, a dynamical companion mass of $30^{+31}_{-12}$~M$_J$, and a mass ratio of $\sim$1.9\%, properties most consistent with low-mass brown dwarfs. However, its mass estimated from luminosity models is a lower $\sim$13.8 $M_{\rm J}$ due to an estimated young age ($\lesssim$ 115 Myr); using a weighted posterior distribution informed by conservative mass constraints from luminosity evolutionary models yields a lower dynamical mass of $23.6_{-7.4}^{+9.1}$~M$_J$ and a mass ratio of $\sim$1.4\%. Analysis of the host star's multi-frequency $\gamma$ Dor-type pulsations, astrometric monitoring of HIP 39017 b, and Gaia Data Release 4 astrometry of the star will clarify the system age and better constrain the mass and orbit of the companion.  This discovery further reinforces the improved efficiency of targeted direct-imaging campaigns informed by long-baseline, precision stellar astrometry.
\end{abstract}



\section{Introduction} \label{sec:intro}

Large, ground-based telescopes assisted by facility -- and, now, specialized \textit{extreme} -- adaptive optics (AO) systems have allowed the imaging of substellar and self-luminous Jovian companions to nearby stars (e.g. \citealt{Marois2008, Lagrange2010, Macintosh2015, Chauvin2018, Currie2023}). Direct imaging is uniquely poised to probe models of planet formation; not only is it sensitive to wide-separation companions in young systems, it also allows both spectral and dynamical orbital analysis for any detected companion. However, large-scale, ``blind'' (or unbiased) surveys have found that planetary and brown dwarf companions that can be imaged with these modern, ground-based systems are rare, with only about a few percent of targeted systems yielding companion detections \citep{Nielsen2019,Vigan2021,Currie2023b}. 

To account for this shortcoming, recent direct imaging surveys have turned toward targeted searches, focusing on stars showing dynamical evidence for a companion. In particular, the 25 yr precision astrometry Hipparcos-Gaia Catalog of Accelerations (HGCA; \citealt{Brandt2018,Brandt2021}) identifies stars displaying acceleration inconsistent with a single-body solution. Such an acceleration could indicate a massive unseen companion, making young, nearby accelerating systems prime targets for imaging. Surveys of accelerating stars have yielded a significantly higher detection rate \citep{Currie2021,Bonavita2022} for low-mass stellar \citep{Steiger2021,Chilcote2021}, brown dwarf \citep{Currie2020a,Bonavita2022,Kuzuhara2022,Franson2023,Li2023}, and even planetary companions of young stars \citep[e.g. HIP 99770 b and AF Lep b;][]{Currie2023, DeRosa2023, Mesa2023, Franson2023b}.

Given the youth of the substellar and planetary companions accessible with direct imaging, they can serve as laboratories for testing and constraining models of their formation and atmospheres \citep{Spiegel2012}. Dynamical mass measurements of imaged companions can provide an independent point of comparison for spectral models. However, due to the large orbits of companions accessible with imaging, the precision of the dynamical masses derived from relative astrometry is often severely limited by small orbital coverage. Selecting imaging targets with existing long-baseline precision astrometric measurements from the HGCA offers the additional benefit of extending orbital coverage for one of the components of the system (e.g. \citealt{BrandtGM2021}).

Age is another crucial factor in constraining the fidelity of atmospheric models for young substellar and planetary companions. These bodies cool with time, with brown dwarfs manifesting this as an apparent evolution to later spectral types. Uncertainties in system ages can have a significant impact on masses retrieved from atmospheric and evolutionary models (e.g. \citealt{MoroMartin2010}). However, if the system is not a member of a coeval stellar association, obtaining precise and reliable independent age constraints can be difficult. While a number of methods to constrain stellar ages from models and empirical relations exist, they may not be applicable to every system (e.g. \citealt{Dantona1984, Mamajek2008}). 

In this work, we report the direct imaging discovery of a massive planet/very-low mass brown dwarf companion $\sim$22 au from the $\gamma$ Doradus ($\gamma$ Dor)type variable accelerating star, HIP~39017, with spectrum resembling that of an L/T transition object. An overview of the host star is provided in Section \ref{ss:systembkgd}. Observations and companion detection are described in Sections \ref{ss:obs} and \ref{ss:detection}. Our spectral and astrometric analyses are presented in Sections \ref{sec:spec} and \ref{sec:orbit}, respectively. In Section \ref{ss:orbitevolpriors}, we revisit the astrometric analysis with posteriors informed by evolutionary models and the HIP 39017 system's likely age. Finally, further discussion is included in section \ref{sec:disc}.

\section{System Properties, Observations, and Data} \label{sec:obs}

\subsection{HIP 39017} \label{ss:systembkgd}

HIP~39017 (HD~65526, V769~Mon) is an A9/F0 star \citep{Henry2011} located at a distance of 65.9~pc \citep{gaiadr3}.  Its spectral type and color-magnitude diagram (CMD) position suggest a mass similar to those of HR 8799 and 51 Eri, $\sim 1.4 - 1.75$~\Msun \citep{Dupuy2022, Sepulveda2022}. \citet{Stassun2019} derive a photometric mass of $1.61 \pm 0.27$~\Msun. 

HIP~39017 is also a $\gamma$ Dor-type variable  \citep{Handler1999, Handler2002, Henry2011}, with three identified variability frequencies with periods ranging from $0.58-0.65$~days \citep{Henry2011}. Its variability was originally ascribed to the presence of an eclipsing binary \citep{Perryman1997,Kazarovets1999}.  However, further analysis, including the identification of multiple variability cycles, resulted in its reclassification as a $\gamma$~Doradus variable \citep{Handler1999, Handler2002, Henry2011}. Thus, HIP 39017 appears to be a single star.  As a $\gamma$ Dor variable, HIP 39017 shares properties with planet-hosting stars HR 8799 and 51 Eri \citep{Marois2010,Moya2010,Macintosh2015,Sepulveda2022b}.

While \cite{Tetzlaff2011} estimate an extremely young age of $\sim 11$~Myr through comparison to several stellar evolutionary models, they assume an A3 spectral type, which is in conflict with the \citeauthor{Henry2011} results.   A comparison with the Banyan-$\Sigma$ tool \citep{Gagne2018} does not identify HIP 39017 as a member of any moving group or young association.   There is currently no evidence that HIP 39017 harbors a luminous debris disk from Wide-field Infrared Survey Explorer, Spitzer, or Herschel results, which are common for early-type stars at such young ages \citep[e.g.][]{Rieke2005}, although available data for HIP 39017 are scarce. 

\begin{figure*}[ht!]
    \centering
    \includegraphics[width=0.7\textwidth,trim=0mm 0mm 0mm 0mm,clip]{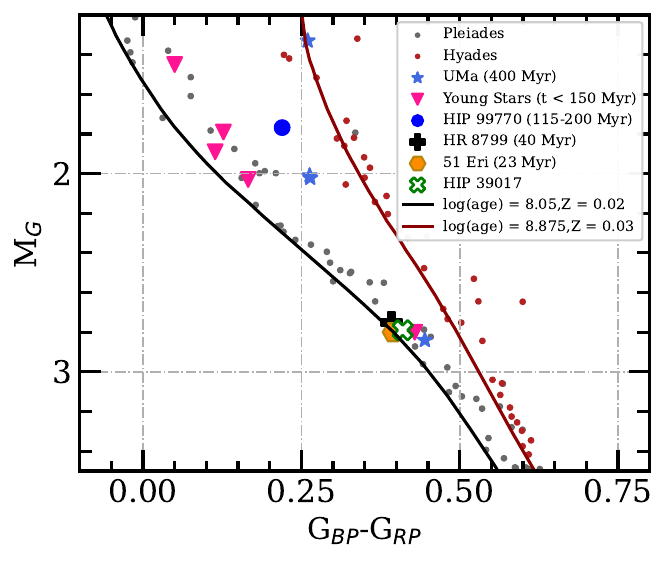}
    \vspace{-0.1in}
    \caption{Color-magnitude diagram drawn from \textit{Gaia} DR3 photometry. We overplot PARSEC isochrones \citep{Bressan2012} appropriate for the Pleiades (t$\sim$115 Myr; dark gray line) and Hyades (t$\sim$750 Myr; red line), add photometry for Pleiades and Hyades members, stars in the 400 Myr old Ursa Majoris association (blue stars; \citealt{Jones2016}), young stars with CHARA-constrained radii and ages (magenta triangles; \citealt{Jones2016}), and stars orbited by directly-imaged planets: 51 Eri, HR 8799, and HIP 99770.  HIP 39017's position lies very close to that of 51 Eri and HR 8799.}
    \vspace{-0.in}
    \label{fig:gaiahrd}
\end{figure*}

On the other hand, the star's \textit{Gaia} CMD position -- $M_{\rm G}$ = 2.79, $B_{\rm P}-R_{\rm P}$ = 0.42 --  is nearly identical to that of the planet-hosting stars HR 8799 (40 Myr; \citealt{Marois2008}) and 51 Eri (23 Myr; \citealt{Macintosh2015}) and consistent with or slightly below the Pleiades sequence, providing evidence of youth (Figure \ref{fig:gaiahrd}). HIP 39017's position is slightly blueward of one Ursa Majoris member (blue star) with an interferometrically-estimated age of 440 Myr \citep{Jones2016} and thus, by implication, younger.  It is significantly bluer than the entire Hyades (750 Myr) locus. Given its CMD position with respect to the Pleiades sequence, an age $\sim 115$~Myr or lower for HIP~39017 is favored. While HIP~39017 is slightly lower metallicity ($\left[\textrm{Fe}/\textrm{H}\right] = -0.26 \pm 0.13$; \citealt{Bruntt2008}) than the Pleiades and Hyades, this difference in metallicity is not sufficient to shift an object of Hyades age or older to the position of HIP~39017 on the CMD. Therefore, we adopt $\sim 115$~Myr as an approximate age of the system, with a conservative upper limit of $\sim700$~Myr for analysis that will later depend on these values (see Section \ref{ss:orbitevolpriors}).\footnote{In principle, HIP 39017's $\gamma$ Dor status could open the door to an asteroseismic age derivation for the star.  However, the path towards deriving an age this way may not apply for all $\gamma$ Dor variables.  While HR~8799 was identified as displaying $\gamma$ Dor-type pulsations \citep{Moya2010}, subsequent analysis found only a single, independent pulsation frequency \citep{Sodor2014}, limiting the application of asteroseismology in deriving the system's age. On the other hand, 51~Eri displays multiple $\gamma$ Dor-type pulsation modes \citep{Sepulveda2022b}, which could allow an asteroseismological age.
}

Despite the star's youth and proximity to the Sun, it was not targeted as a part of major, recent high-contrast imaging surveys conducted with the \textit{Spectro-Polarimetric High-contrast Exoplanet REsearch} or \textit{Gemini Planet Imager} instruments \citep{Beuzit2019,Macintosh2014} and has not been observed with any other AO instrument on any 8-10 m telescope nor with the \textit{Hubble Space Telescope}.   However, the early Data Release 3 (eDR3) version of the HGCA \citep{Brandt2021} reveals that HIP~39017 has a statistically significant astrometric acceleration ($5.5\sigma$), plausibly due to an unseen substellar companion.   HIP~39017 is bright enough ($R \sim 6.8$~mag; \citealt{Monet2002}) to achieve high performance of ground-based extreme-AO.   Thus, we targeted it as a part of the SCExAO high-contrast imaging survey of nearby, young accelerating stars \citep{Currie2021,Currie2023}.

\begin{deluxetable*}{llllllllll}
     \tablewidth{0pt}
    \tablecaption{HIP 39017 Observing Log\label{obslog_hip39017}}
    \tablehead{
    \colhead{UT Date} &
    \colhead{Instrument} &
    \colhead{$\theta_{\rm{v}}$} & 
    \colhead{Bandpass} &
    \colhead{Bandpass Wavelength} &
    \colhead{$t_{\rm exp}$} & 
    \colhead{$N_{\rm exp}$} 
    & $\Delta$PA
    \\ 
    \colhead{} &  
    \colhead{} &  
    \colhead{(arcsec)} & 
    \colhead{} & 
    \colhead{($\mu m$)} & 
    \colhead{(s)} & 
    \colhead{}&
    \colhead{(deg)} &
     \colhead{} 
    }
    \startdata
    \centering
    2022-02-21& SCExAO/CHARIS & 0.60--0.80 & Broadband/JHK & 1.16--2.37 & 60.48 & 113 & 66.2  \\
    2022-03-21 & Keck/PyWFS+NIRC2 &  0.83-1.33 & $L^{\prime}$ & 3.78 & 27.0 & 100 & 43.6    \\
    2022-03-25& SCExAO/CHARIS\tablenotemark{a} & 0.7--0.9 & Broadband/JHK & 18 & 5.3 & 5 & 4.1  \\
    2022-12-31& SCExAO/CHARIS & 0.60?\tablenotemark{b} & Broadband/JHK & 1.16--2.37 & 30.98 & 137 & 55.3 \\
    2023-02-04& Keck/NIRC2 & 0.76-1.10 & $H$ & 1.63 & 20 & 5 & 0.4
    \enddata
    \tablecomments{       
    $\theta_{\rm{v}}$ represents the characteristic seeing measurements from the Canada France Hawaii Telescope seeing monitor.
    The integration time of each exposure, the number of exposures used in our analysis, and the total variation of parallactic angle in each sequence are represented by $t_{\rm exp}$, $N_{\rm exp}$, and $\Delta PA$, respectively.}
        \tablenotetext{a}{These CHARIS data had the star offsetted by 0\farcs{}5 to bring the background star within the field of view}
      \tablenotetext{b}{The CFHT seeing monitor did not record values, but the on-sky PSF quality was consistent with median to slightly better-than-median seeing.}
\end{deluxetable*}

   \begin{figure*}[ht!]
    \centering
    \includegraphics[width=0.325\textwidth,trim=0mm 0mm 0mm 0mm,clip]{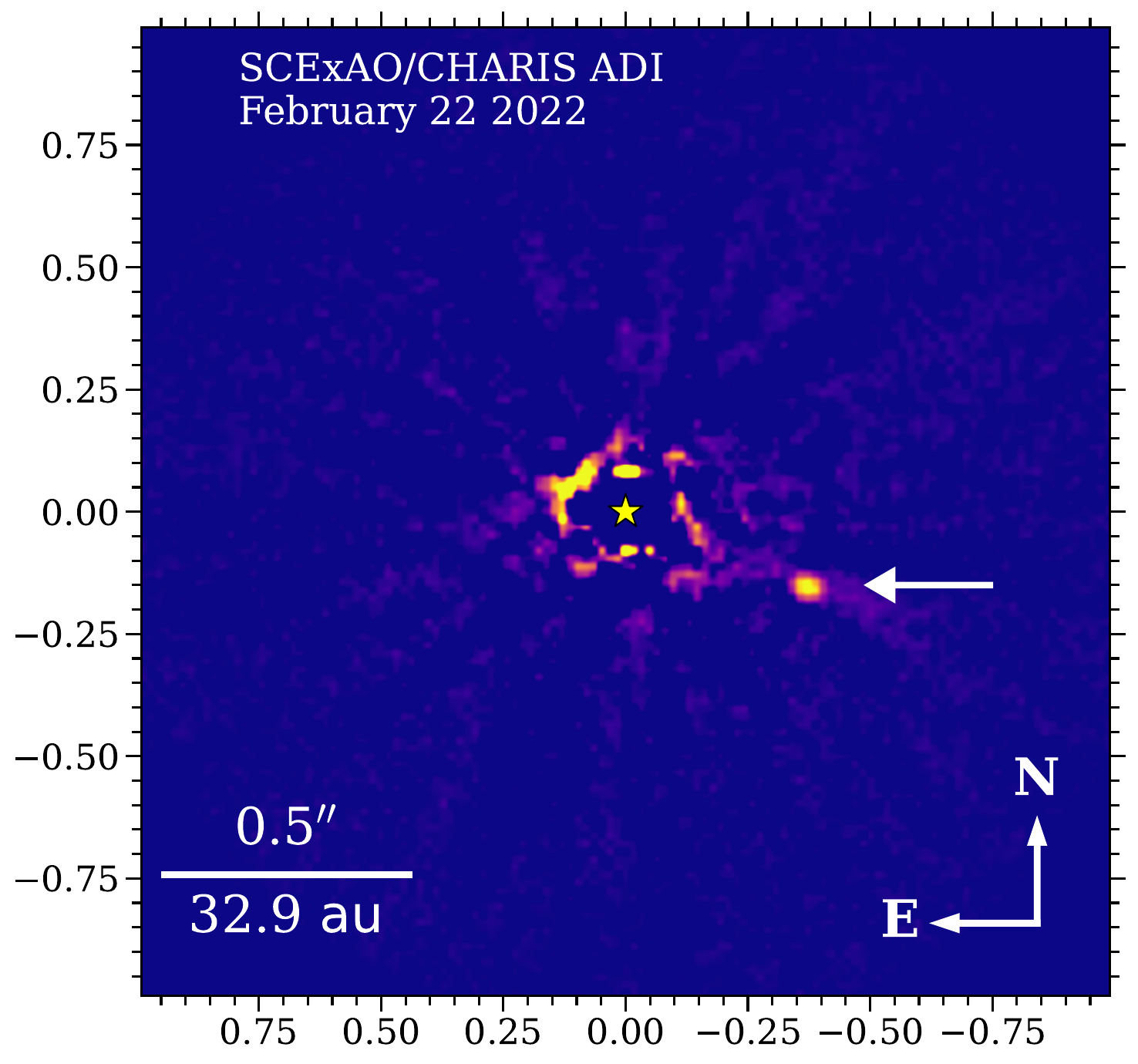}
      \includegraphics[width=0.325\textwidth,trim=0mm 0mm 0mm 0mm,clip]{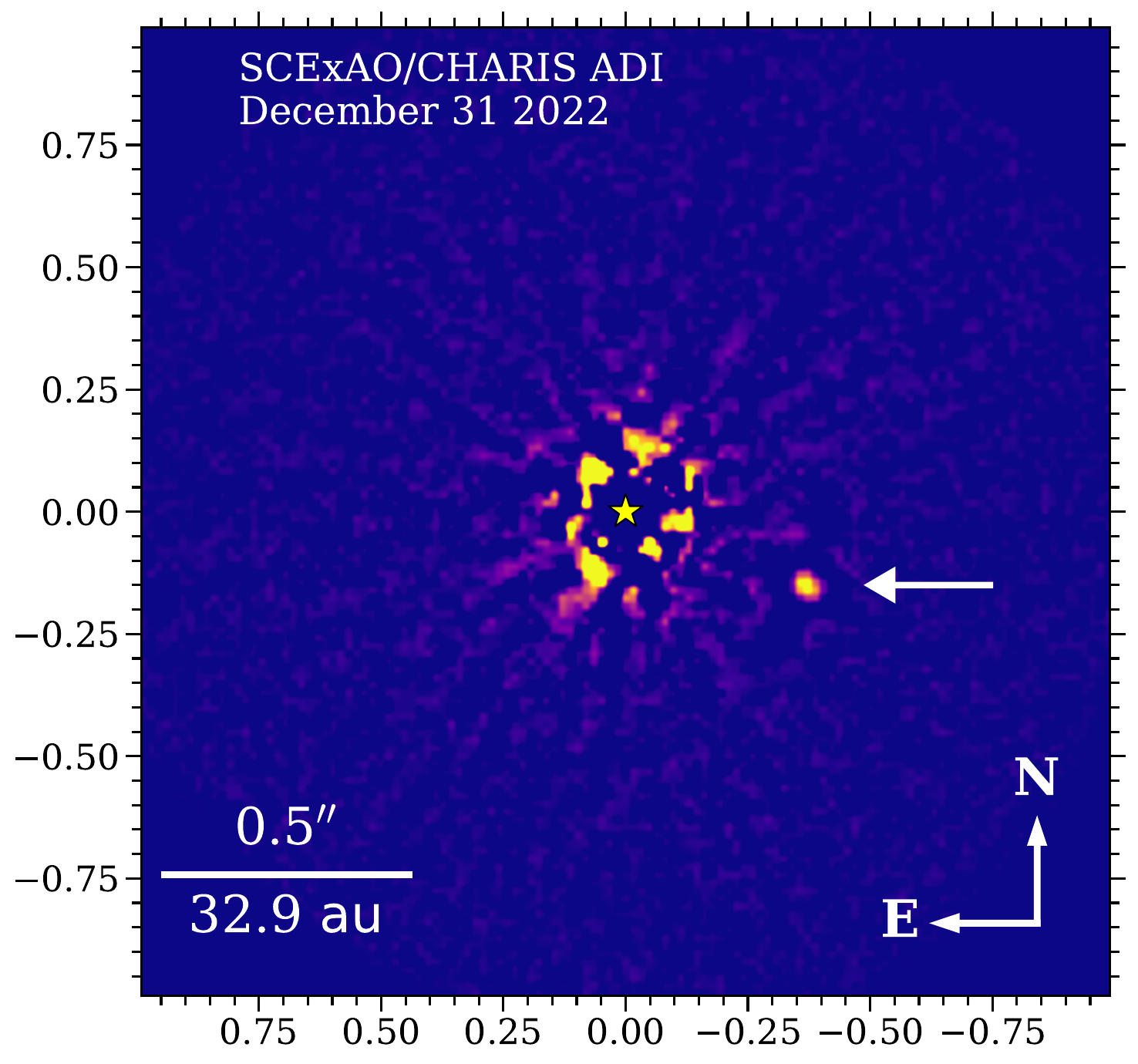}
        \includegraphics[width=0.325\textwidth,trim=0mm 0mm 0mm 0mm,clip]{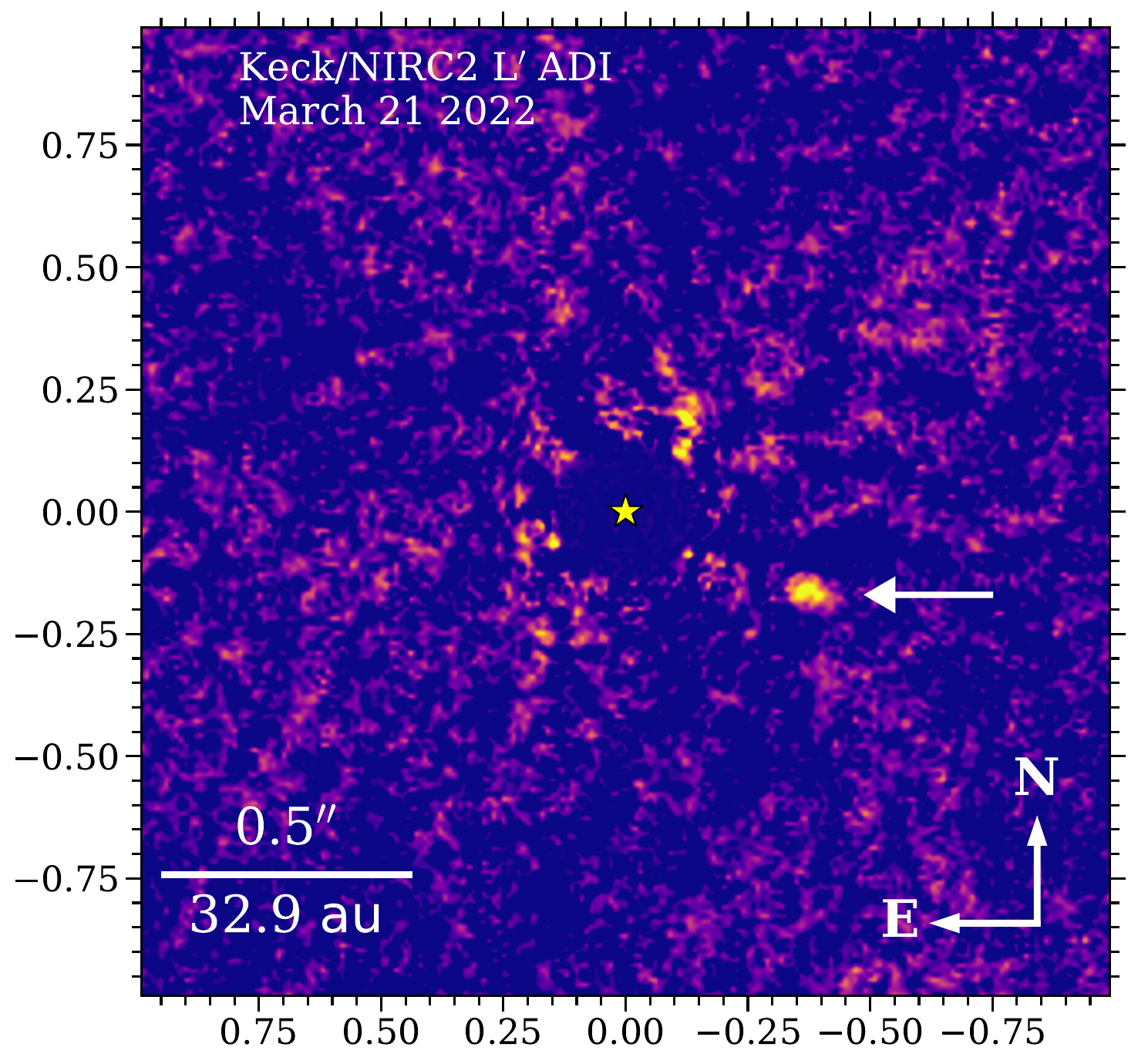}
    \vspace{-0.1in}
    \caption{Detections of HIP 39017 b (denoted with arrows) with SCExAO/ CHARIS IFS data at 1.1--2.4 $\mu m$ (left, middle panels) and Keck/NIRC2 in the $L^{\prime}$ filter at 3.78 $\mu m$ (right).   }
    \vspace{-0.in}
    \label{fig:hip39017image}
\end{figure*}

\subsection{High-Contrast Imaging Observations}\label{ss:obs}

We conducted AO-assisted, high-contrast imaging observations of HIP 39017 using the Subaru and Keck II telescopes on Maunakea on five nights between February 2022 and March 2023  (Table \ref{obslog_hip39017}).    For the Subaru observations, the facility AO system, AO188 \citep{Hayano_2008_AO188}, provided a first stage correction of atmospheric turbulence, and SCExAO then delivered a higher order correction of residual wavefront errors \citep{Jovanovic2015,Currie2020b}; Keck II observations used either the near-IR Pyramid wavefront sensor \citep[PyWFS][]{Bond2020} (March 2022) or the facility Shack-Hartmann wavefront sensor (other data sets).  The Subaru and Keck observations utilized a Lyot or vector vortex coronagraph, respectively, to suppress residual stellar halo light.  For Subaru, sharpened starlight was then sent to the CHARIS integral field spectrograph (IFS; \citealt{Groff2016}), while the NIRC2 camera recorded Keck images.

All data were obtained in pupil tracking/\textit{angular differential imaging} (ADI) mode \citep{Marois2006}.
CHARIS data were obtained in low resolving power ($\mathcal{R} \sim$ 18) spectroscopic mode to obtain wide wavelength coverage (1.1--2.4 $\mu m$), while NIRC2 data were taken in the $L^{\rm \prime}$-band broadband filter.   Satellite spots produced by a modulation of SCExAO's deformable mirror provided spectrophotometric calibration for our CHARIS data and were used to register the star to a common center \citep{Jovanovic2015b, Sahoo2020}.  For NIRC2, we calibrated the photometric measurements by obtaining unsaturated images of the central star before and after the main coronagraphic sequence. For the March 2022 CHARIS data, we offsetted the star from the image plane center by 0\farcs{}5 to follow up a wide-separation candidate companion seen in the first Keck/NIRC2 epoch that is outside the nominal CHARIS field of view and was shown to be a background star (see Appendix).

We used data cubes extracted from the raw CHARIS data by the Automated Data Extraction, Processing, and Tracking System for CHARIS (ADEPTS; \citealt{Tobin2020, Tobin2022}), which utilizes the CHARIS Data Reduction Pipeline (DRP; \citealt{Brandt2017}).
We performed subsequent processing steps on the extracted CHARIS cubes -- sky subtraction, image registration, spectrophotometric calibration, point-spread function subtraction, and spectral extraction --  using the CHARIS Data Processing Pipeline \citep{Currie2020b}. For spectrophotometric calibration, we assumed an F0V model atmosphere from the Kurucz library \citep{Castelli2003} with the star's Two Micron All Sky Survey (2MASS) photometry \citep{Skrutskie_2006_2MASS,doi-2MASS}.   For the February 2022 and December 2022 data, we subtracted the PSF using A-LOCI \citep{Currie2012,Currie2015} in combination with ADI only.  For the shallower March 2022 data focused on an outer candidate, we simply subtracted the radial intensity profile of each slice of each cube before rotating all cubes to the north and median-combining.

We processed the NIRC2 $L^{\prime}$ images with the \texttt{VIP} package from \citet{Gomez2017} and the $H$-band data using the general-use pipeline from \citet{Currie2011,Currie2014_HR8799}. For pre-processing, we correct for pixel sensitivity variations through flat-fielding and dark subtraction in our science images. Cosmic ray artifacts are removed using the $\texttt{lacosmic}$ Python package \citep{vanDokkum2001}, and geometric distortion is addressed combining solutions from \citet{Service2016} and the script from \citet{Yelda_2010} for the NIRC2 camera's narrow-field mode. The post-alignment data cube has a centering stability of less than 0.5 mas, which we also take into account in our final astrometry. For  $L^{\prime}$-band data post-processing, we applied the annular PCA algorithm in \texttt{VIP} with six principle components to subtract stellar PSF, following the methods outlined in \citet{Li2023}. For the $H$-band data, we applied a radial profile subtraction to each image before combining north-rotated images.

\subsection{Detection of HIP 39017 b}\label{ss:detection}

Figure \ref{fig:hip39017image} shows the HIP 39017 images obtained from two CHARIS data sets (February 2022 and December 2022) and one NIRC2 data set (March 2022).  Each data set reveals a faint companion about $\approx$0\farcs{}4 southwest from the star.  The companion, hereafter HIP 39017 b, is detected with signal-to-noise ratios (SNRs) of 6--24 \footnote{The SNRs were calculated following the approach in \citet{Currie2011}, considering the finite-element correction of \citet{Mawet2014}.}.  The December 2022 CHARIS detection is the highest fidelity one, as HIP 39017 b is visible in each channel. The signal-to-noise and relative astrometry for the companion in each epoch are listed in Table \ref{tbl:det_hip39017}.

To correct our HIP 39017 b SCExAO/CHARIS spectrophotometry and astrometry for biasing due to processing, we carried out forward modeling as in \citet{Currie2018}.   HIP 39017 b's throughput ranged from 27\% to 80\%, depending on the data set and channel.   Our forward-modeling analysis indicates minimal astrometric biasing ($\sim$ 0.05 pixels in both x and y).   To estimate astrometric errors, we imputed the planet forward-model at the same approximate angular separation but 1000 different azimuthal angles, compared the imputed and recovered astrometry, and computed the scatter in these values \citep[e.g.][]{Kuzuhara2022}.   Based on these analyses, we estimate an intrinsic centroiding uncertainty of $\sim$0.1 pixels in both coordinates, which we add in quadrature with an assumed stellar centroiding uncertainty of 0.25 pixels and converted to values in polar coordinates.  

For Keck/NIRC2 $L^{\prime}$ data, we opted for a simpler method of inserting negative copies of HIP 39017 b over a range of separations and with a range of brightnesses until the companion's signal is fully nulled.  While the SCExAO/CHARIS detections have a higher SNR, the instrument's centroiding uncertainty coupled with the larger north position angle offset result in an uncertainty comparable to NIRC2's in the companion's position angle. 

\begin{deluxetable*}{llllllllll}
     \tablewidth{0pt}
    \tablecaption{HIP 39017 b Astrometry\label{tbl:det_hip39017}}
    \tablehead{
    \colhead{UT Date} &
    \colhead{Instrument\tablenotemark{a}} &
    \colhead{Bandpass} &
    \colhead{S/N} & 
    \colhead{$\rho$} &
    \colhead{PA} 
    \\ 
    \colhead{} &  
    \colhead{} &  
    \colhead{} & 
     \colhead{} &
     \colhead{(mas)} &
     \colhead{(deg)} 
    }
    \startdata
    2022-02-21& SCExAO/CHARIS & Broadband/$JHK$ & 19.3 & 404.1 $\pm$ 4.3 & 247.351 $\pm$ 0.616 \\
    2022-03-21 & Keck/PyWFS+NIRC2 & $L^{\prime}$ &6.1 & 401.1 $\pm$ 16.0 & 247.556 $\pm$ 0.650 \\
    2022-12-31& SCExAO/CHARIS & Broadband/$JHK$ & 24.4 & 400.4 $\pm$ 4.3 & 247.698 $\pm$ 0.622 \\
    \enddata
      \tablenotetext{a}{The wavelength range for CHARIS is 1.16--2.37 $\mu$m, while the $L^{\rm{\prime}}$-filter's central wavelength for NIRC2 is 3.78 $\mu$m.}
    \tablecomments{
    S/N represents the object's signal-to-noise ratio, $\rho$ represents the companion.}
\end{deluxetable*}

\section{Infrared Spectrum and Photometry} \label{sec:spec}

\begin{figure}
    \centering
       \vspace{-0.0in}
    \includegraphics[width=0.45\textwidth,trim=0mm 0mm 0mm 0mm,clip]{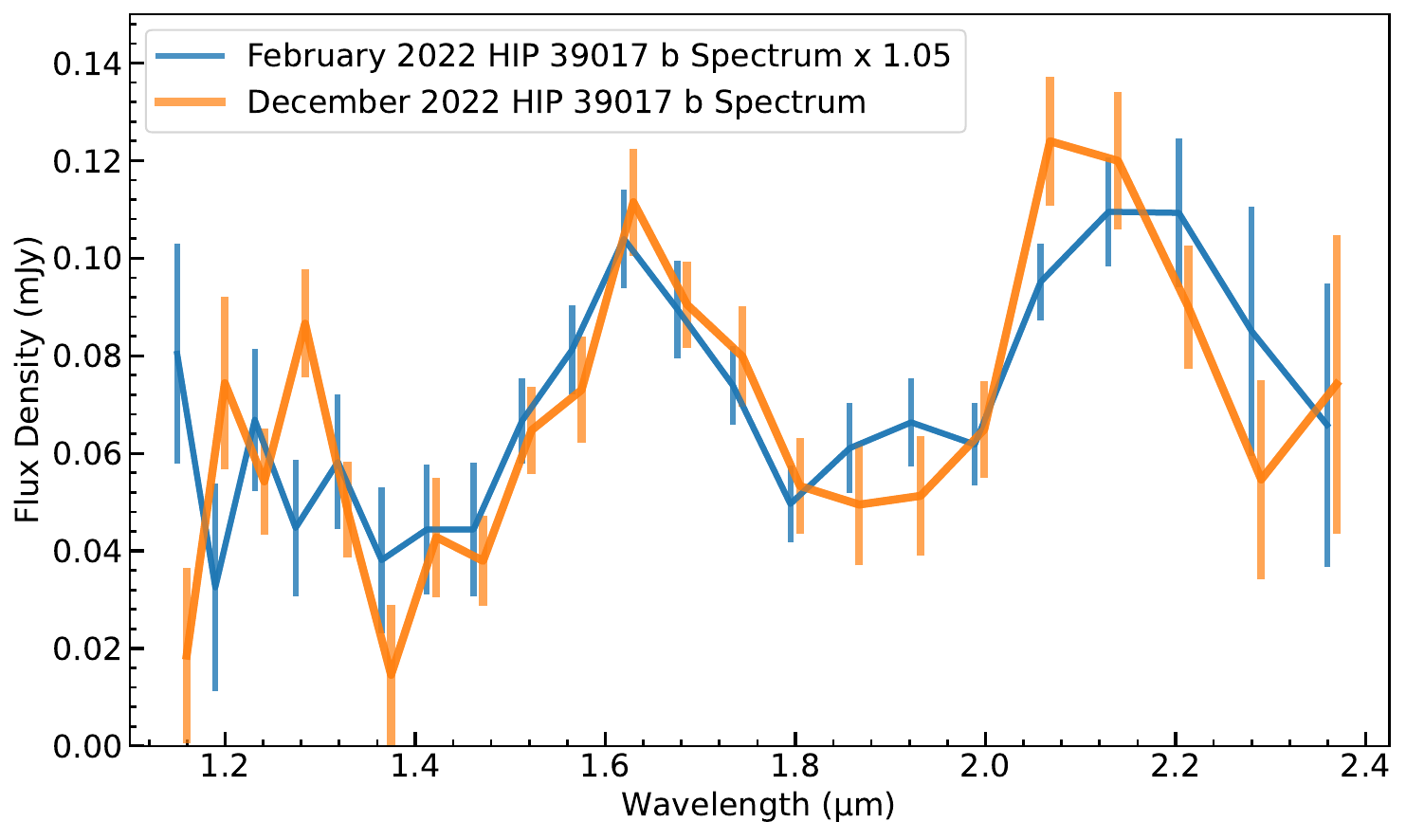}\\
    \vspace{-0.1in}
    \caption{CHARIS spectrum of HIP 39017 b extracted from February 2022 (royal blue) and December 2022 (orange) data.  Because of its slightly higher SNR and lower levels of residual speckle contamination in J band, we focus our analysis on the December spectrum. }
    \vspace{-0.in}
    \label{fig:hip39017spectrum}
\end{figure}

Figure \ref{fig:hip39017spectrum} shows the CHARIS spectrum for HIP 39017 b extracted from both epochs.   The February 2022 spectrum is much noisier in J band due to greater residual speckle noise contamination.   However, the spectra agree within errors in 18 of the 22 channels, including all but one channel in the H and K passbands: the February data appear to have a systematic offset of $\sim$ 5-10\%, consistent with slight variations in the absolute spectrophotometric calibration.   We estimate HIP 39017 b's broadband photometry  in the major Mauna Kea Observatories passbands from December 2022 CHARIS data to be $J$ = 18.31 $\pm$ 0.20, $H$ = 17.75 $\pm$ 0.13, and $K_{\rm s}$ = 17.02 $\pm$ 0.14, consistent with a substellar object at the L/T transition \citep[e.g.][]{DupuyLiu2012,Currie2023}\footnote{Photometry derived from the February 2022 data formally agrees within errors: $J$ = 18.91 $\pm$ 0.41, $H$ = 17.81 $\pm$ 0.13, and $K_{\rm s}$ = 17.12 $\pm$ 0.13.}. 

We derive a companion-to-host contrast of $\Delta L^{\prime} = 9.93\pm0.07$  magnitudes for the NIRC2 $L^{\prime}$ band data.  Using the intrinsic colors for main sequence stars from \citet{Kenyon1995} and \citet{PecautMamajek2013}, we adopt a  primary star brightness of $m_{\rm L^{\prime}, \star} = 6.19\pm0.044$ mag, the same as its W1 magnitude \citep{Cutri_2012}, as the two are equivalent in the Rayleigh-Jeans tail of A stars.  HIP~39017~b's brightness in $L^{\prime}$ is then $16.12\pm0.18$ mag. 

\begin{figure*}
    \centering
    \includegraphics[width=\textwidth]
    {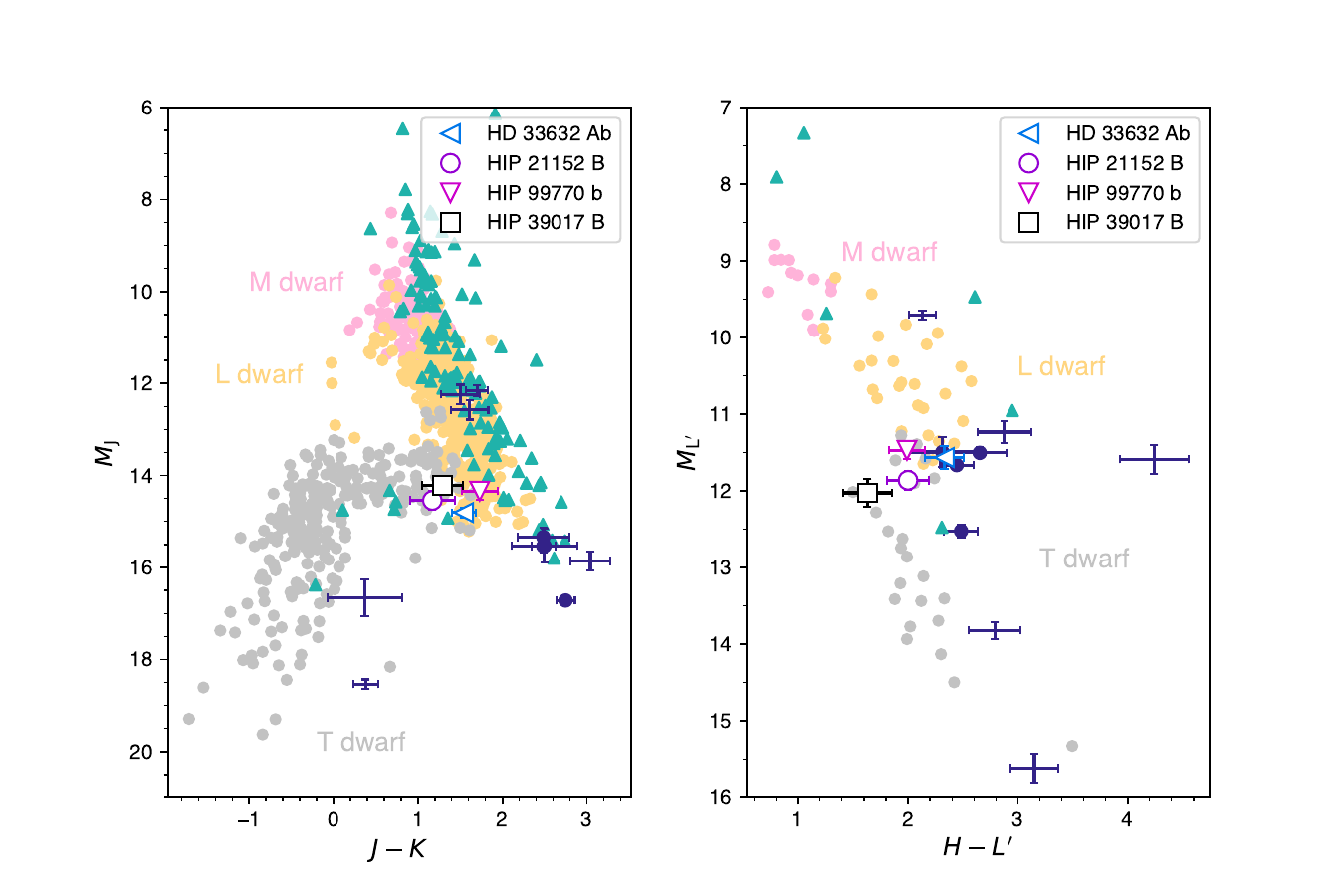}
    \caption{Color-magnitude diagram in the Mauna Kea Observatories passbands for substellar objects as in \citet{Kuzuhara2022} but adding photometry for HIP~99770b \citep{Currie2023} and HIP~39017b (this work).  
    Teal triangles denote young or low surface-gravity MLT dwarfs \citep{DupuyLiu2012, Leggett2010, Best2020} using distances from \citet{Best2020}. HIP 99770 b \citep{Currie2023} and substellar companions HIP~21152B \citep{Kuzuhara2022} and HD~33632~Ab \citep{Currie2020a} are included for comparison, as well as a selection of other planets and brown dwarf companions denoted by indigo error bars: GJ~504~b \citep{Kuzuhara2013,Janson2013}, HD~95086b \citep{DeRosa2016}, 51~Eri~b \citep{Best2020,Rajan2017}, 2M1027~b \citep{Best2020,DupuyLiu2012} $\beta$~Pic~b \citep{Best2020,DupuyLiu2012}, 1RXS J160929.1-210524B \citep{Best2020,DupuyLiu2012}, and $\kappa$~And~b \citep{Best2020,DupuyLiu2012},
    all using distances from \citet{BailerJones2021}. Indigo error bars with filled circles are the HR~8799 planets \citep{Marois2008, Metchev2009, Skemer2014, Currie2014_HR8799, Zurlo2016, Skrutskie_2006_2MASS, doi-2MASS}.  
    }
    \label{fig:cmd}
\end{figure*}

Figure \ref{fig:cmd} compares HIP 39017 b's broadband photometry to measurements for directly-imaged planets, and directly-imaged very low-mass brown dwarfs, and field brown dwarfs.   The companion's position in both $J$/$J$-$K$ and $L^{\prime}$/$H$-$L^{\prime}$ CMDs overlaps with objects at the L/T transition \citep{Saumon2008}.  HIP 39017 b's colors also show a strong resemblance to those for directly-imaged companions around accelerating stars HIP 99770 b,  HD 33632 Ab, and especially HIP 21152 B.   

\subsection{Comparison to Empirical Spectra} \label{ss:empspec}

Next, we compared the CHARIS spectrum from the December 2022 data set to the archive of published L and T dwarfs from the SpeX Prism Spectral Libraries \citep{Burgasser2014}. 

The template spectra were first resampled to linearly-spaced bins using the \texttt{FluxConservingResampler} in Astropy's \texttt{specutils} package \citep{specutils,FluxConservingResampler}, before being smoothed and resampled down to the same $R$ as the CHARIS spectrum. Each template spectrum was then fit by minimizing 
\begin{equation}
    \chi^2 = \sum_k ( f_{\nu,k} - \alpha F_{\nu,k} )^T C_k^{-1} ( f_{\nu,k} - \alpha F_{\nu,k} ),
    \label{eq:chi2calc}
\end{equation}
where $F_{\nu,k}$ and $f_{\nu,k}$ are the flux of the template and target spectra, respectively, in the $k^{th}$ bin, and $\alpha$ is the scaling factor. The covariance matrix, $C$, includes the uncertainties from both the template and target spectra. The covariance for the target SCExAO/CHARIS spectrum is derived and parameterized following \citet{Greco2016}, with the December spectrum best fit by $A_{\rho}=0.0675$, $A_{\lambda}=0.0525$, $A_{\delta}=0.88$, $\sigma_{\rho}=0.8175$, and $\sigma_{\lambda}=2.8$. In each fitting iteration, the uncertainty of the template spectrum, scaled by the amplitude $\alpha$ for that iteration, is added in quadrature to the diagonal of the covariance matrix for the target \citep{Currie2018}.

\begin{figure}
    \centering
    \includegraphics[width=0.47\textwidth]{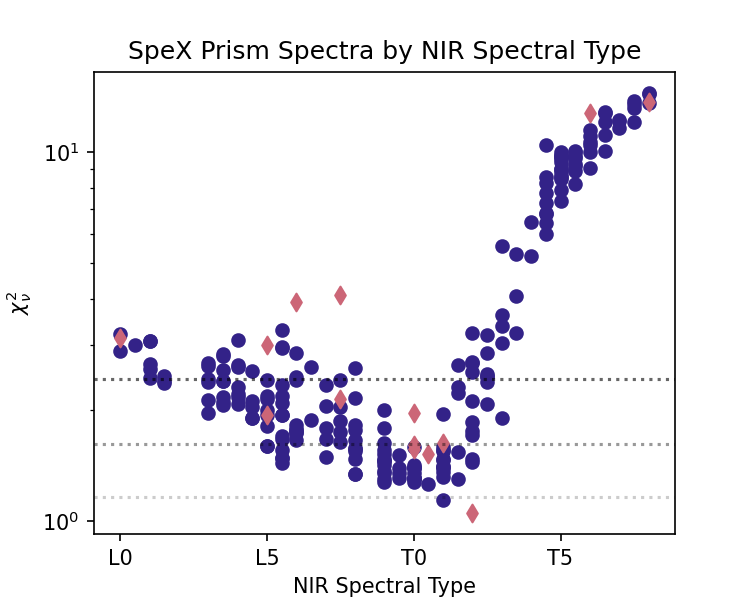}
    \caption{$\chi_{\nu}^2$ of each SpeX Prism template spectrum fit to the HIP 39017b CHARIS spectrum versus its NIR spectral type as given in the SpeX Prism library. Template spectra with `peculiar' NIR spectral type are shown as diamonds. Dotted lines indicate the $\chi_{\nu}^2$ values that correspond to the 1, 2, and 3 $\sigma$ confidence intervals, from bottom to top.}
    \label{fig:spex_chi2}
\end{figure}

\begin{figure}
    \centering
    \includegraphics[width=0.44\textwidth]{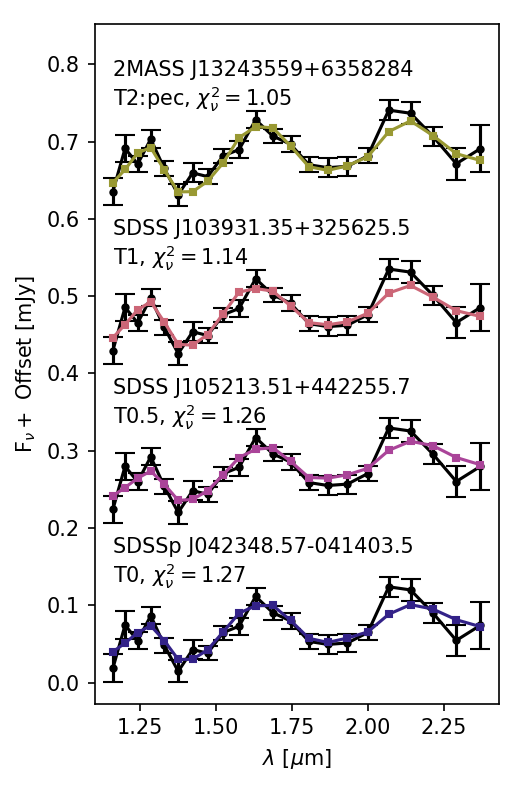}
    \caption{The four smoothed and binned, best fit SpeX Prism spectra (in color) compared to the HIP 39017b Broadband CHARIS spectrum from December 2022 (black), plotted with vertical offset for visualization. Uncertainties on the SpeX Prism library spectra are plotted, but errorbars appear small compared to the marker size, due to CHARIS's larger spectral resolution. NIR spectral types are as given in the SpeX Prism library \citep{Looper2007, Burgasser2010, Burgasser2008, Burgasser2004}.}
    \label{fig:spex_spectra}
\end{figure}

Figure \ref{fig:spex_chi2} shows the reduced-$\chi^2$ achieved for each fitted spectrum as a function of its near-infrared (NIR) spectral type, and Figure \ref{fig:spex_spectra} shows the four best-fit spectra compared to that of HIP 39017b. Within $1\sigma$, the SpeX Prism sources that best fit the HIP 39017b spectrum indicate that it is $\sim$T1-2, though a closer look at those best-fit sources may be more accurately described as a peculiar T$1-2$ dwarf. 

The best-fit source to the HIP 39017~b $JHK$ SCExAO/CHARIS spectrum is 2MASS J13243553+6358281 \citep{Looper2007}. Classified at the time as a peculiar T2 dwarf, spectral matching analyses postulated that its unusually red colors \citep{Metchev2008} could be a result of an unresolved binary \citep{Looper2007,Burgasser2010}, though a more recent identification of this source as a member of the $\sim 150$~Myr old AB Doradus group suggests that its redness may instead be ascribed to its young age \citep{Gagne2018}. HIP~39017's young age is certainly consistent with the appearance of a redder spectrum for HIP~39017b, like that seen in 2MASS J13243553+6358281. 

The only other SpeX Prism library source with a $\chi_{\nu}^2$ fit within $1\sigma$ is SDSS J103931.35+325625.5 (2MASS J10393137+3256263; \citealt{Burgasser2010}). While its NIR spectrum was closest to that of a T1 dwarf, it was also classified as a possible unresolved binary \citep{Burgasser2010}. 

Although $L^{\prime}$ photometry was not included in this fit, we briefly compare the measured or estimated $H-L^{\prime}$ color of the four best-fit SpeX library sources in Figure \ref{fig:spex_spectra} to that of HIP~39017b, $H-L^{\prime}=1.63 \pm 0.22$. Of the four sources, only SDSSp J042348.57-041403.5 has measured MKO $L^{\prime}$ photometry \citep{Leggett2002}. For the remaining three, we use the photometric transformation relation from \citet{Leggett2019} to estimate their $L\prime$ from Spitzer $3.6$ and $4.5 \mu$m photometry \citep{Kirkpatrick2021, SpitzerSourceList2021, Leggett2007}, combined with measured MKO $H$-band photometry \citep{Zhang2021, Chiu2006}. We find that the measured and derived $H-L^{\prime}$ colors of these four sources are somewhat redder than HIP~39017b, with values between $2.1 - 2.5$~mag. However, a comparison of the absolute magnitudes is complicated by the unknown binarity of most of these library sources. In fact, SDSSp J042348.57-041403.5 has been resolved into a close L$6.5+$T$2$ binary \citep{DupuyLiu2017}.

\begin{deluxetable*}{lcccccccc}
    \tablewidth{0pt}
    \tablecaption{Best Fit Spectral Models}
    \tablehead{
    \colhead{Model Grid} &
    \colhead{$T_{eff}$} &
    \colhead{$\log g$} &
    \colhead{C/O} & 
    \colhead{[M/H]} &
    \colhead{clouds} &
    \colhead{$r$CH$_4$\tablenotemark{a}} &
    \colhead{$\chi_{\nu}^2$} &
    \colhead{$R$} \\
    \colhead{} &
    \colhead{(K)} &
    \colhead{} &
    \colhead{} & 
    \colhead{} &
    \colhead{} &
    \colhead{} &
    \colhead{} &
    \colhead{(R$_{J}$)}
    }
    \startdata
    SONORA BOBCAT & 1700 & 4.0 & 1.0 & 0.0\tablenotemark{b} & NC\tablenotemark{b} & No\tablenotemark{b} & 2.82 & 0.51$_{-0.02}^{+0.02}$ \\
    SONORA DIAMONDBACK & 1300 & 4.0 & 1.0\tablenotemark{b} & 0.0 & f3 & No\tablenotemark{b} & 1.10 & 0.87$_{-0.03}^{+0.03}$  \\
    LACY/BURROWS 99770 & 1500 & 4.5 & 1.0\tablenotemark{b} & 0.0\tablenotemark{b} & AEE100 & No & 1.59 & 0.69$_{-0.02}^{+0.02}$ \\
    \enddata
      \tablenotetext{a}{Whether or not reduced CH$_4$ abundances were reduced by a factor of 10. Grids with reduced CH$_4$ only available for the Lacy/Burrows 99770 models.}
      \tablenotetext{b}{Only value of this parameter available in the model grid.}
    \tablecomments{Companion radius, $R$, is derived from the fitted model flux using the Gaia DR3 system distance ($d=65.8835 \pm 0.0916$~pc; \citealt{gaiadr3}).}\label{tbl:spec_fits}
\end{deluxetable*}
\subsection{Comparison to Atmospheric Model Spectra} \label{ss:modspec}

We compare the HIP~39017b spectrum from the December 2022 CHARIS observations along with the Keck/NIRC2 $L^{\prime}$ photometry to three model atmosphere grids: the Sonora Bobcat models \citep{Marley2021}, the Sonora Diamondback models (\citediamondback), and the Lacy/Burrows models generated for the analysis of HIP~99770b \citep{Currie2023}. 

Fitting followed a similar procedure as the empirical spectral fits described in Section \ref{ss:empspec}, with a few caveats. Due to the lack of uncertainty arising from the model spectral points, the covariance matrix, $C_k^{-1}$, used when calculating the fit is just that of the measured CHARIS $JHK$ spectrum. The calculation of $\chi^2$ from Equation \ref{eq:chi2calc} also includes the additional photometric term, $(f_{\nu,L'} - \alpha F_{\nu,L'} )^2 / \sigma_{F\nu,L'}^2$. Each model spectrum is fit to the observed spectrum by varying the scaling parameter for the entire model spectrum, $\alpha$. Additional parameters for each model spectrum are as defined in the pre-generated model grid. A summary of the best fit model parameters and their $\chi_{\nu}^2$ for each model grid is given in Table \ref{tbl:spec_fits}.

The Sonora models are a family of self-consistent atmospheric models for L, T, and Y dwarfs and self-luminous, young planets \citep{Marley2021, Karalidi2021}. Bobcat, the first of the Sonora models released, invokes equilibrium chemistry and cloudless atmospheres \citep{Marley2021}. The released spectral grid \citep{SonoraBobcatData} spans $T_{eff} = 200 - 2400$~K and $\log g = 3.0 - 5.5$ for solar metallicity ([M/H]$= 0$) and C/O, with extensions of this grid for C/O values of 0.5 and 1.5 available for $\log g = 5.0$. 

The upcoming Sonora Diamondback models (\citediamondback) include clouds. The spectral grid (C. Morley 2023, private communication) covers $T_{eff} = 900 - 2400$K, $\log g = 3.5-5.5$, and [M/H]$= \lbrace -0.5, 0.0, 0.5 \rbrace$ for six different realizations of clouds: $f_{sed} = \lbrace 1, 2, 3, 4, 8 \rbrace$ and no clouds (``$nc$''), where $f_{sed}$ is the sedimentation efficiency (\citealt{Ackerman2001}; labeled therein as $f_{rain}$). Clouds with more efficient sedimentation (higher $f_{sed}$) are able to condense more effectively, leading to a lower optical depth \citep{Ackerman2001}. 

To investigate an alternative cloud parameterization, we also compared the HIP~39017b spectrum to the Lacy/Burrows model grid derived for comparison to HIP~99770b \citep{Currie2023}. The Lacy/Burrows models expand on those of \citet{Burrows2006} with updated molecular line lists and absorption cross sections, as well as non-equilibrium carbon chemistry \citep{Currie2023}. While less expansive in their coverage of the $T_{eff}$ ($\sim 1100 - 1600$K) and $\log g$ (4.0, 4.5, 5.0) parameter space, they do include the ($T_{eff}$, $\log g$) that provided the best fit in the cloudy Sonora Diamondback models. Further, cloud parameterization is differentiated into the modal size of the atmospheric dust particles, (in microns, \citealt{Currie2023}) and cloud structure:AE, AEE, and E, in order of increasing sharpness of the cutoff at the top of the clouds \citep{Madhusudhan2011}. As these cloud models have the same interior profile, a more gradual cutoff with height means the presence of more cloud material and therefore a higher total optical depth \citep{Madhusudhan2011}. 

The Lacy/Burrows model spectra from \citet{Currie2023} contain grids of AE- and AEE-type clouds with a modal dust size of $a_0 = 100 \mu$m with default methane abundances, as well as grids of AEE-type, $a_0 = 100\mu$m clouds and E-type, $60\mu$m clouds with CH$_4$ abundance reduced by a factor of 10.

The Sonora Bobcat model that best fits the HIP~39017b spectrum has $T_{eff} = 1700$K, $\log g =4.0$, and C/O$=1.0$, albeit with a significantly higher $\chi_{\nu}^2$ (2.82) than is achieved by either of the other models. Notably, the best fit Bobcat models all fall short of the high peak flux observed around $\lambda \sim 2.07 - 2.14 \mu$m in the HIP~39017b spectrum (see Appendix \ref{apx:dec-bobcat}).

\begin{figure*}[t]
    \centering
    \includegraphics[width=0.92\textwidth]{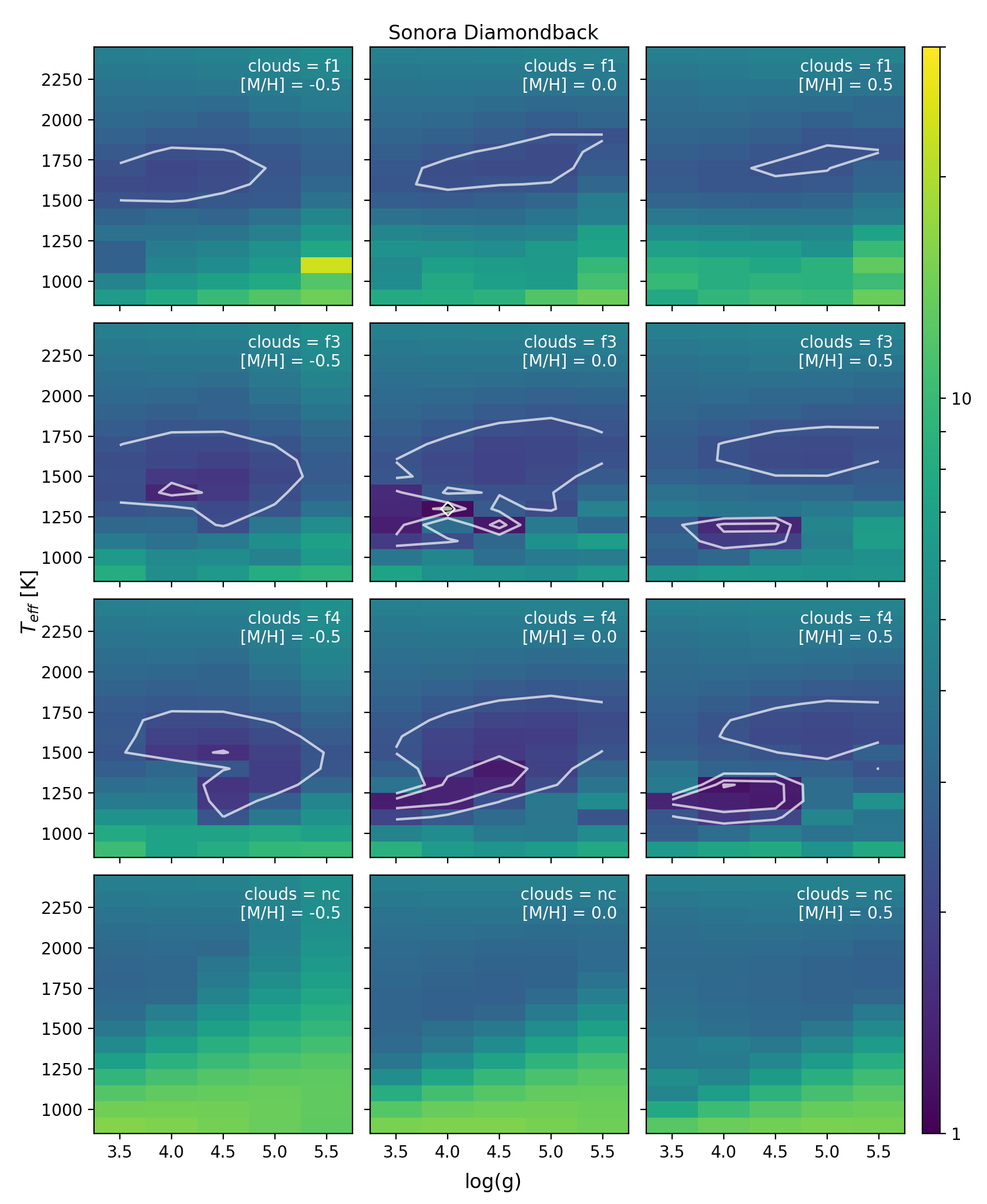}
    \caption{Maps of the $\chi_{\nu}^2$ derived from fitting the HIP~39017b spectrum to the Sonora Diamondback (\citediamondback) model spectrum grids as a function of model $\log g$ and $T_{eff}$. Columns of subfigures show the three available metallicities: $[M/H] = -0.5$ \textit{(left column)}, 0.0 \textit{(middle column)}, and $+0.5$ \textit{(right column)}. Rows of subfigures show four of the six different cloud prescriptions: $f_{sed} = 1 $ \textit{(top row)}, 3 \textit{(second row)}, 4 \textit{(third row)}, and no clouds \textit{(fourth row)}.}
    \label{fig:diamondback_chi2map}
\end{figure*}

\begin{figure*}
    \centering
    \includegraphics[width=0.48\textwidth]{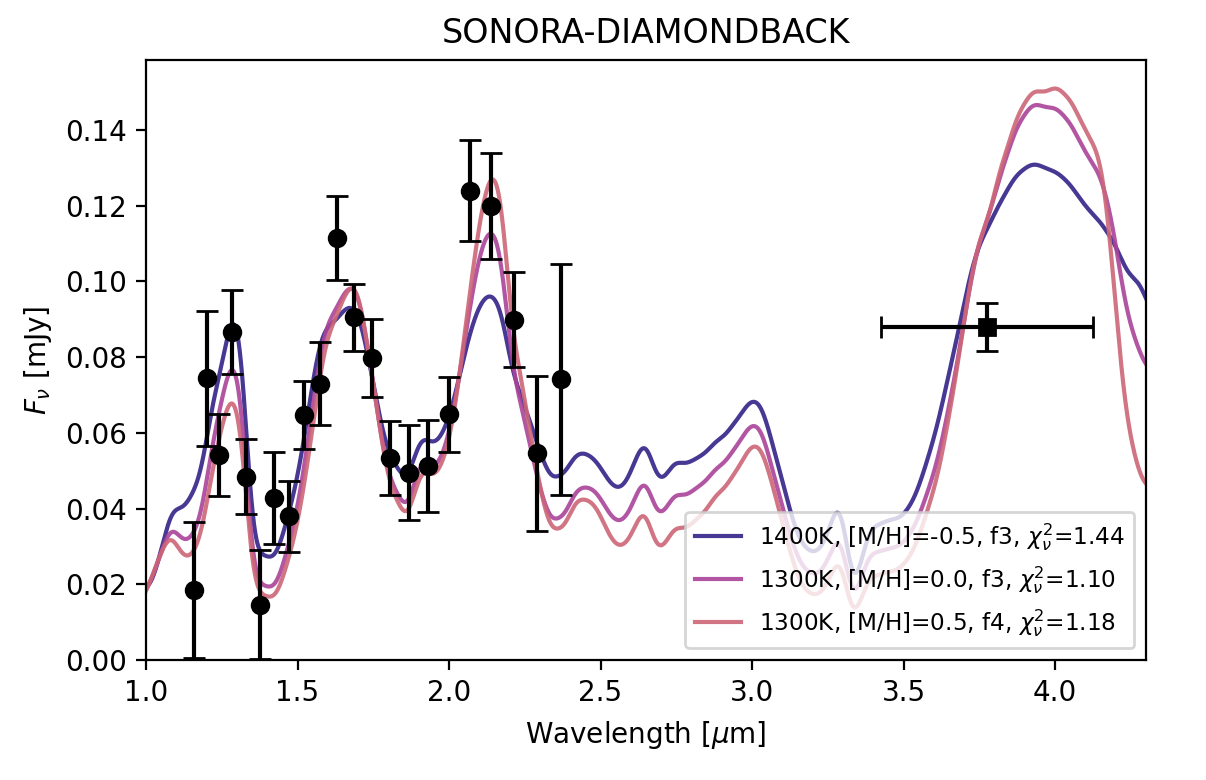}
    \includegraphics[width=0.48\textwidth]{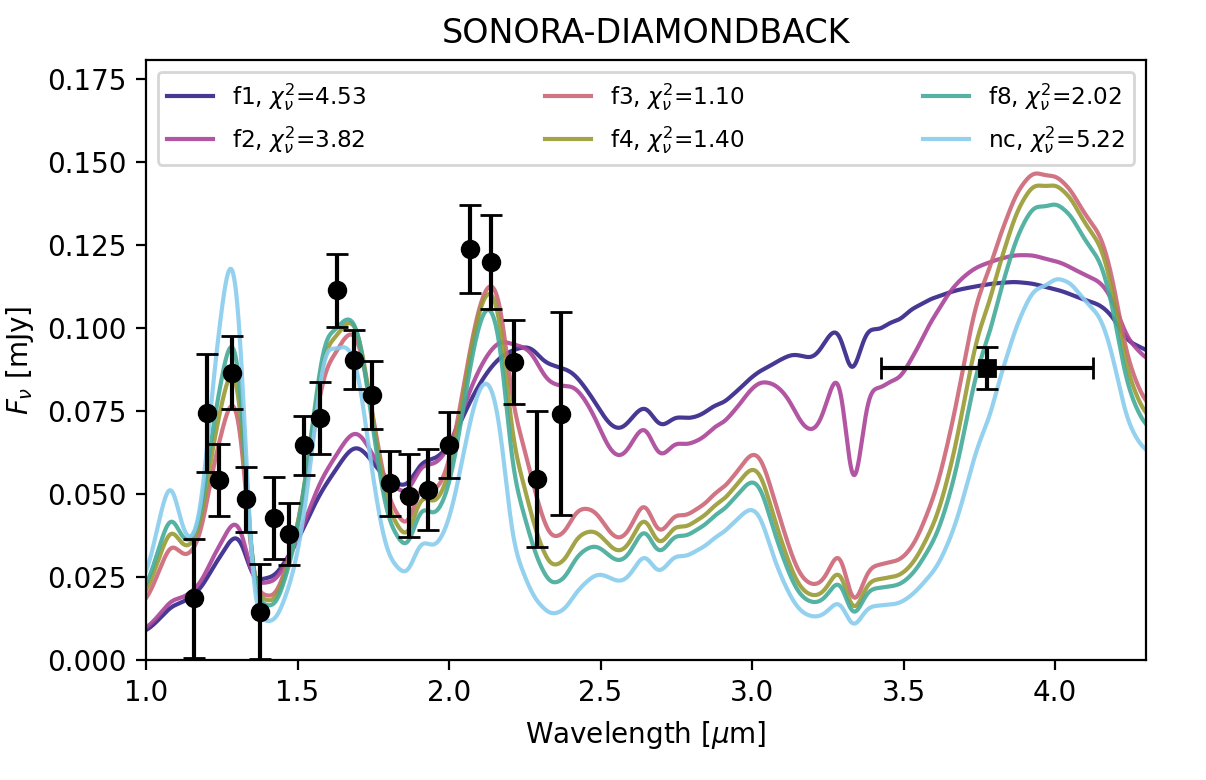}
    \caption{Select Sonora Diamondback (\citediamondback) model spectra, smoothed to $R=50$ at $2.65\mu$m, overplotted with the HIP~39017b observations. \textit{Left:} The best fit Sonora Diamondback spectra with each metallicity provided in the spectral grid. The $T_{eff}$, [M/H], and $\chi_{\nu}^2$ for each model are given in the legend. All three best-fit spectra have $\log g = 4.0$. \textit{Right:} Model spectra with the same $T_{eff}$ (1300~K), $log(g)$ (4.0), and [M/H] (0.0) as the best fit spectrum, but showing variation with the cloud prescription. Each spectrum's $f_{sed}$ value (or `nc' for no clouds) and its corresponding $\chi_{\nu}^2$ are given in the legend.}
    \label{fig:diamondback_spec}
\end{figure*}

The results of the Sonora Diamondback fits are shown in Figures \ref{fig:diamondback_chi2map} and \ref{fig:diamondback_spec}. The addition of clouds with the Sonora Diamondback models provide a smaller $\chi_{\nu}^2$. The HIP~39017b spectrum is best fit by the 1300~K, $\log g =4.0$, [M/H]$=0.0$ model with $f_{sed}=3$ ($\chi_{\nu}^2 = 1.099$). The only other model spectrum that falls within $1\sigma$ of this minimum has the same $T_{eff}$ and $\log g$, but [M/H]$=+0.5$ and $f_{sed} = 4$.

A comparison of these two model spectra (as well as the best fit model spectrum with [M/H]$=-0.5$) can be seen in the first panel of Figure \ref{fig:diamondback_spec}. These two best fit models both show a much more accurate fit to the peak flux observed at $\sim 2.07 - 2.14 \mu$m than the cloudless models, as well as a much higher peak $\sim 3.8 \mu$m. 

The second panel of Figure \ref{fig:diamondback_spec} shows how different model spectra with the same $T_{eff}$ (1300~K), $\log g$ (4.0), and [M/H] (0.0) as the best fit model compare to the observations when the cloud $f_{sed}$ varies. As can be seen in the figure, clouds with $f_{sed} \sim 3-8$ are necessary to achieve the heightened $K$-band peak and slightly damped $J$-band peak detected in the HIP~39017b CHARIS spectrum from 2022 December under the Diamondback model. Clouds with $f_{sed} \sim 1-2$, on the other hand, induce too much reddening of the spectrum compared to our observations of HIP~39017b, flattening the $JHK$ peaks significantly.

\begin{figure*}
    \centering
    \includegraphics[width=0.54\textwidth]{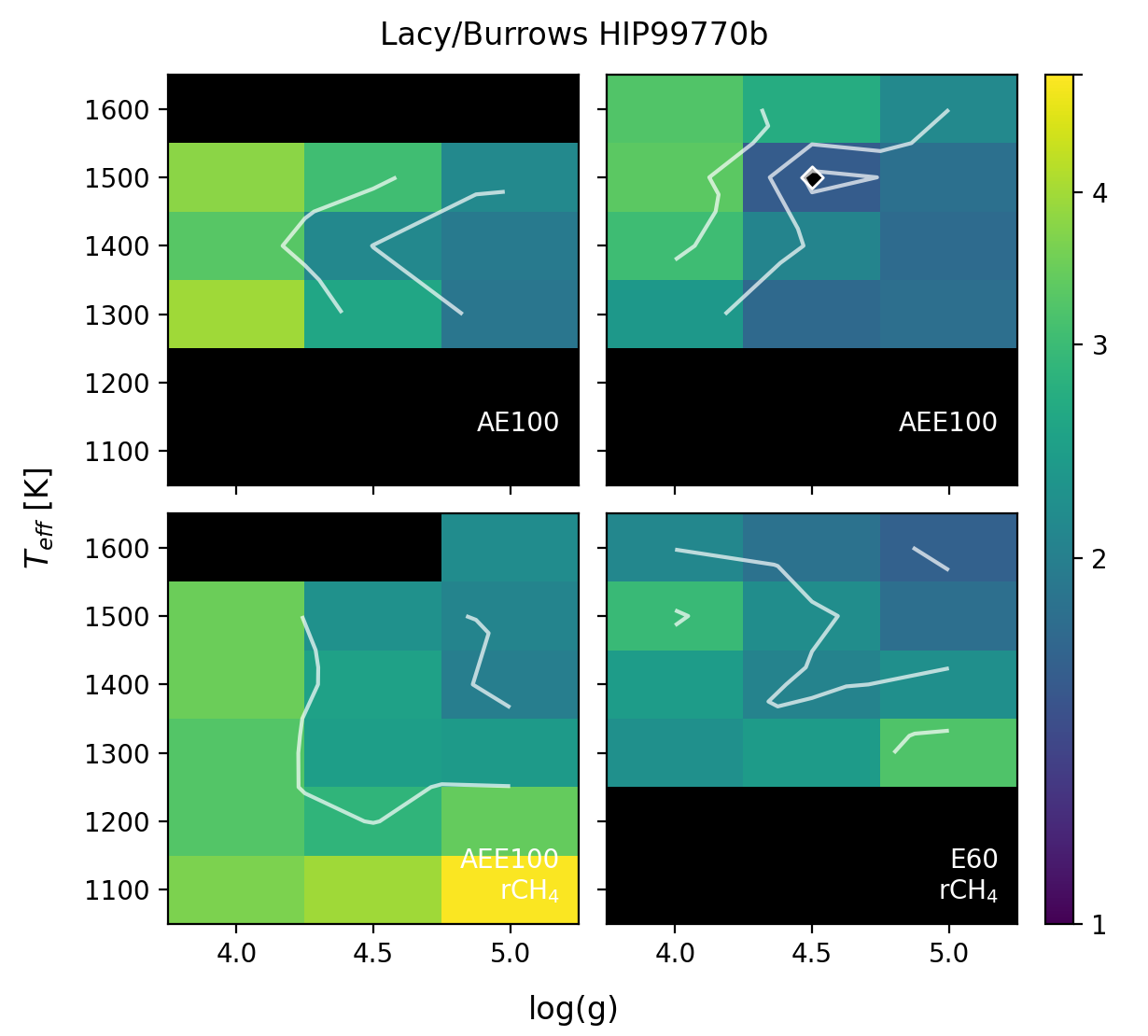}
    \includegraphics[width=0.54\textwidth]{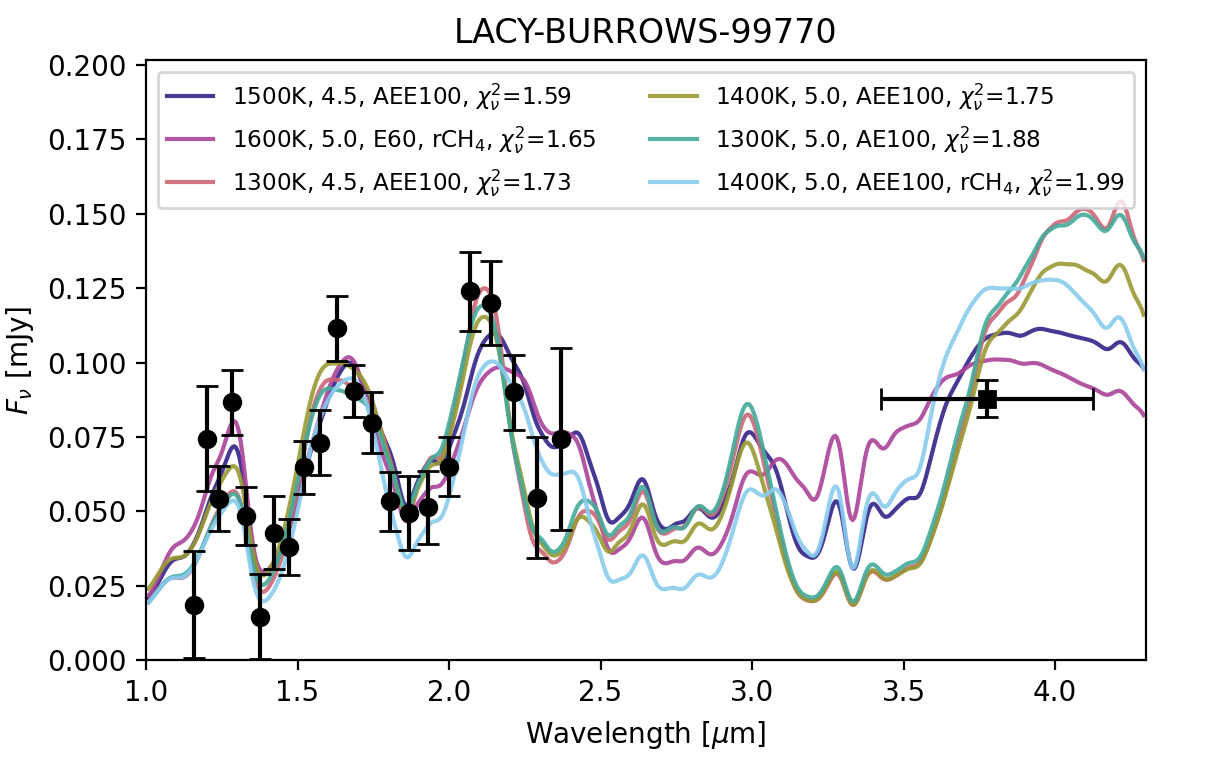}
    \caption{The fits results of the Lacy/Burrows models \citep{Currie2023} to the December 2022 SCExAO/CHARIS spectrum and the NIRC2 $L^{\prime}$ photometric point of HIP~39017b. (\textit{Top}) $\chi_{\nu}^2$ fits for each available model spectrum as a function of $T_{eff}$ and $\log g$ as above. Cloud parameterization is labeled for each subfigure (AE, AEE, or E, as well as modal dust size, in microns), with `rCH$_4$' indicating grids derived with reduced methane. (\textit{Bottom}) Six of the Lacy/Burrows model spectra, smoothed to $R=50$ at $2.65\mu$m and overplotted on the observed HIP~39017b spectrum used for fitting. The legend gives the $T_{eff}$, $\log g$, cloud prescription, whether they have reduced CH$_4$, and its $\chi_{\nu}^2$ fit to the observations. The first four spectra shown are the four Lacy/Burrows spectra with the lowest $\chi_{\nu}^2$, while the last two are those with the lowest $\chi_{\nu}^2$ of the remaining two cloud prescription/methane abundance grids.}
    \label{fig:lacyburrows}
\end{figure*}

The results of our fit to the Lacy/Burrows models are shown in Figure \ref{fig:lacyburrows}. The best-fit Lacy/Burrows model from the HIP~99770b model grid for HIP~39017b is 1500~K, $\log g = 4.5$ with solar CH$_4$ and AEE-type clouds with $100\mu$m dust. Compared to the Sonora Diamondback model fits above, this represents a slightly warmer and higher gravity minimum than found for the extensive Diamondback grids (1300~K, $\log g = 4.0$, $f_{sed}=3$). While no significant minimum is seen at 1300~K and $\log g = 4.0$ in the Lacy/Burrows grid, this point is at the boundary of the parameter space covered for all but one of the cloud grids. 

However, as can be seen from the fine coverage of the Diamondback grids, it is not unusual for the location of the minimum $\chi_{\nu}^2$ to vary with cloud strength; in fact, local minima at higher $T_{eff} \sim 1500$~K and $\log g \sim 4.0$ can be seen in several of the Diamondback grids in Figure \ref{fig:diamondback_chi2map} (e.g. $f_{sed}=4$ and [M/H]$=0$). Therefore, the difference in parameters at which the best fit is achieved may be due in part to a difference in parameter space coverage.

The substantial improvement in the best-fit $\chi_{\nu}^2$ between the cloudless Sonora Bobcat models and the cloudy Sonora Diamondback and Lacy/Burrows models implies that clouds are not only present on HIP~39017b, but are also having a significant impact on its NIR spectrum. While the best-fit models in all three grids analyzed here have similar $\log g$, [M/H], and C/O, the $T_{eff}$ of the best-fit model is lower for the cloudy models than the cloudless grid. Among the Sonora grids, the inclusion of clouds causes the best fit $T_{eff}$ to shift from 1700~K in Bobcat to 1300~K in Diamondback, the latter of which is perhaps more consistent with its apparent position near the L/T transition in Section \ref{ss:empspec} \citep{Kirkpatrick2005}. 

In addition, the radius implied for the best-fit spectrum in each suite of model grids is shown in Table \ref{tbl:spec_fits}, as derived from the scaled model flux and the Gaia Data Release 3 (DR3) distance for the system ($d=65.8835 \pm 0.0916$~pc; \citealt{gaiadr3}). In calculating $R$, we include an estimated 7\% systematic uncertainty in the amplitude of the measured flux. This combines a $\sim 5\%$ uncertainty from the intrinsic variability of brown dwarf spectra \citep{Cruz2018} and a $\sim 5 \%$ calibration uncertainty in the flux of the CHARIS $JHK$ spectral points due to variability in the satellite spot flux ratios \citep{Wang2022}.

Scaled model spectrum flux often yields unphysically small radii for brown dwarfs (e.g. \citealt{Zalesky2019, Gonzales2020, Hood2023, Zhang2023}), which is generally attributed to uncertainty or absent physical mechanisms in the models \citep{Kitzmann2020,Lueber2022}. Indeed, for the three model spectrum grids analyzed here, $R$ ranges from $0.51 - 0.87$~R$_J$. For comparison, $\log g = 4.0$ ($4.5$) for a $30$M$_J$ object would correspond to a radius of $\sim 2.7$R$_J$ ($1.5$R$_J$), while even a $13$M$_J$ object would correspond to a radius of $\sim 1.8$R$_J$ ($1.0$R$_J$). The two cloudy models do seem to underestimate the radius (or equivalently, overestimate the flux) less than the cloudless Bobcat model, as expected with the improved $\chi_{\nu}^2$ of their spectral fits.

\section{Astrometric Analysis} \label{sec:orbit}
\subsection{HIP 39017 b Is Not a Background Star}
\begin{figure}
    \centering
       \vspace{-0.0in}
     \includegraphics[width=0.44\textwidth,trim=0mm 0mm 0mm 0mm,clip]{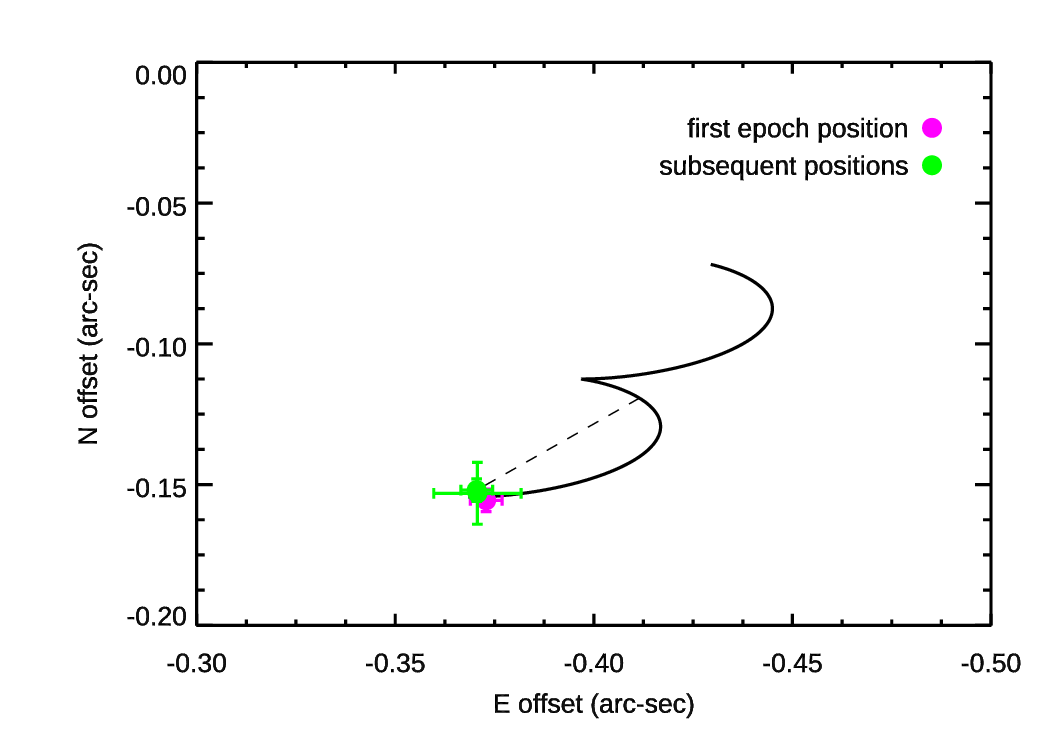}
          \includegraphics[width=0.44\textwidth,trim=0mm 0mm 0mm 0mm,clip]{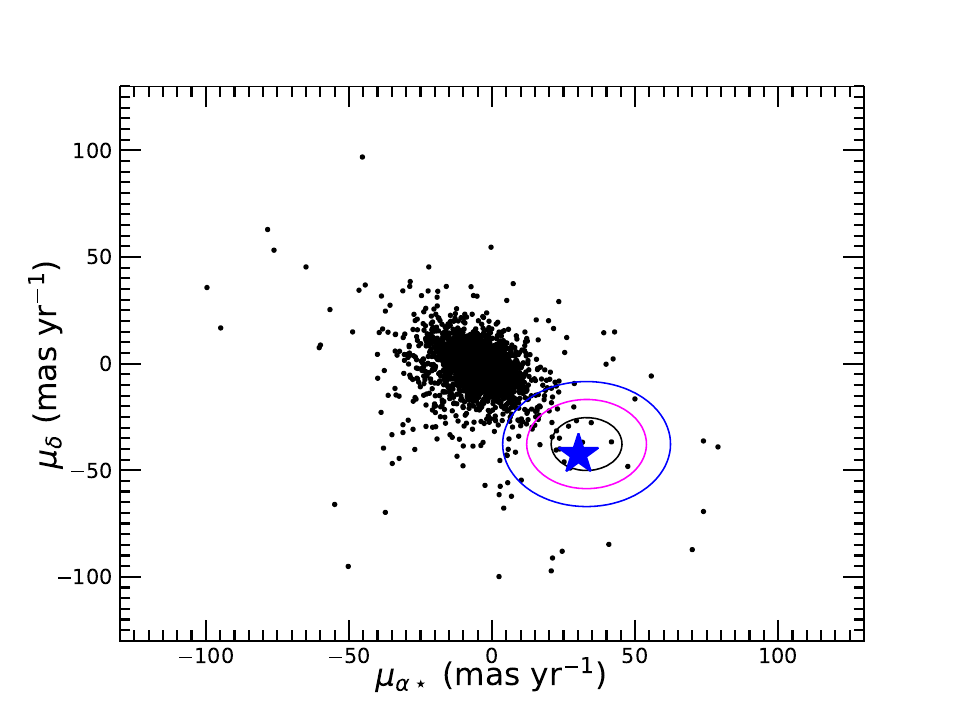}
         \vspace{-0.1in}
    \caption{\textit{Top:} Proper motion analysis for HIP 39017 b compared to the motion of a background star with zero proper motion. \textit{Bottom:} Proper motion analysis for a synthesized Besancon population compared to the 1, 3, and 5-$\sigma$ contours for background stars needed to match HIP 39017 b's relative astrometry.}
    \vspace{-0.in}
    \label{fig:propmot}
\end{figure}

Figure \ref{fig:propmot} (top panel) compares HIP 39017 b's astrometry to the motion expected for a background star.  We rule out a stationary background star at a $>$12-$\sigma$ level.  In comparison, a background object seen at $\approx$1\farcs{}7 in some data sets follows the expected background object track (see Appendix \ref{apx:bkgd-src}).

Following \citet{Nielsen2017} and \citet{Currie2023b}, we compute the relative probability of HIP 39017 b being a background star ($P(BG)P(\rho|BG)P(\nu|\rho)$) with a non-zero common proper motion versus a comoving companion $P(comp)\times~P(\rho|comp)$.   We adopt a Besancon model, modeling the background star distribution over a square-degree area with distances ranging from 0 kpc to 1 kpc ($\pi$ $>$ 1 mas).  Our model considers 3,085 stars as faint as $H$ $\sim$ 20.6, or $\sim$2$\times$10$^{-6}$ times the brightness of the primary (roughly our contrast limit at HIP 39017 b's separation). 

About 55 stars match HIP 39017 b's kinematics to within the 5-$\sigma$ level (Figure \ref{fig:propmot}, bottom panel) for $P(\nu|\rho)$ = 55/3085.     For a 2-$\sigma$ range over our entire set of observations of $\rho$ $\sim$ 0\farcs{}385--0\farcs{}417 and assuming we would detect HIP 39017 b between 0\farcs{}25 and 1\farcs{}1 from the star, we find $P(BG)P(\rho|BG)$ $\sim$ 1.9$\times$10$^{-5}$ for a combined background probability of 3.4$\times$10$^{-7}$.  For a frequency of 0.8\% for 13 $M_{\rm J}$--80 $M_{\rm J}$ companions and mass and semimajor axis distributions of $dN_{\rm comp}/dM$ $\propto$ $M^{-0.47}$ and $dN_{\rm comp}/da$ $\propto$ $a^{-0.65}$, we derive a planet/brown dwarf probability of $\sim$8.6$\times$10$^{-5}$.  Thus, the relative probability of a background star versus a planet/brown dwarf is $\sim$ 0.004: a bound companion is more probable at the $\approx$ 3-$\sigma$ level.   

As found in \citet{Currie2023b}, the true probability of a background star is lower if HIP 39017 b's spectral type is considered.    For a flat detection limit of 2$\times$10$^{-6}$ from 0\farcs{}25 to 1\farcs{}1 and absolute $H$-band magnitudes from \citet{PecautMamajek2013}, a T0 dwarf must be closer than $\approx$ 250 $pc$.  Adopting the space density of late L/early T dwarfs of $\approx$2$\times$10$^{-3}$ pc$^{-3}$, we estimate a contamination rate of $\approx$10$^{-6}$ on the CHARIS field and $\approx$5$\times$10$^{-5}$ for our entire survey ($\approx$60 stars) to date.   

\subsection{Dynamical Modeling}
Using the relative astrometry derived in section \ref{ss:detection} along with proper motion measurements of the stellar host from Gaia DR3 (+30.22 mas yr$^{-1}$ in R.A., and -42.53 mas yr$^{-1}$ in decl.), we have verified a common proper motion between the host star HIP~39017~A and HIP~39017~b. Our observations also revealed a wider separation candidate companion near the edge of the field of view in all datasets. Through detailed analysis of our direct imaging observations and relative astrometry, this peripheral object has been conclusively classified as a background star demonstrating proper motions of the same magnitudes but directly contrary to the movement of the HIP~39017~Ab system (see Appendix~\ref{apx:bkgd-src} for a detailed explanation).

The absolute astrometry from HGCA (eDR3; \citet{Brandt2021}) complements the newly acquired relative astrometry in facilitating the orbital retrieval of HIP~39017~A. HIP~39017~A is a $\gamma$  Doradus variable known to exhibit large stellar pulsations with flux variations of 0.6-6$\%$ or radial velocity (RV) variations in the range of 2-4 km s$^{-1}$. The only four epochs of existing RV data \citep{DeCat2006, Henry2011} show significant fluctuations within the range of 2-4 km s$^{-1}$, rendering these RV data not useful for orbit analysis. In the absence of RV data, we conduct orbital analysis solely based on relative astrometry of HIP 39017 b from high-contrast imaging and absolute astrometry of HIP 39017 A from the HGCA.
\subsubsection{Methodology}\label{sec:dynammethod}
We use the open-source Bayesian orbit modeling code \texttt{orvara} \citep{Brandtorvara} to infer the orbital parameters of HIP~39017~b within an  Markov Chain Monte Carlo (MCMC) framework. \texttt{orvara} utilizes the \texttt{emcee} \citep{Foreman-Mackey_2013} ensemble sampler for parallel tempering MCMC (PT-MCMC). This technique involves multiple parallel chains with varying temperatures, facilitating efficient sampling near the minimum $\chi^2$. Chains at different temperatures swap positions periodically to enhance information sharing and sampling efficiency. Typically, the coldest chain in PT-MCMC efficiently explores the parameter space and finds the global optimum likelihood. The parameter space \texttt{orvara} samples is nine-dimensional, including semi-major axis, eccentricity, reference epoch longitude ($\lambda_{\text{ref}}$), inclination, argument of periastron ($\omega$), and longitude of ascending node ($\Omega$)), system masses ($M_{pri}$ and $M_{sec}$), and an RV jitter term. The code enhances computational efficiency by analytically marginalizing auxiliary model parameters such as RV zero-points, parallax, and barycentric proper motion.

We analyze the orbital properties and mass of HIP 39017 b in several steps as follows.
First, we perform a simple joint \texttt{orvara} fit to the full datasets of relative
and absolute astrometry to derive posterior distributions for orbital parameters, the primary mass, and companion mass.  Over the course of one year, HIP 39017 b moves little between subsequent CHARIS and NIRC2 data sets, in contrast to the behavior for other recently imaged companions like HIP 99770 b and HIP 21152~B \citep{Currie2023,Kuzuhara2022}.   Thus, we expect that constraints on at least some HIP 39017 b parameters may be poorer.  Therefore, we then  compare HIP 39017 b's luminosity and estimated age ($\lesssim$~115 Myr) to predictions for hot-start luminosity evolution models.   These models give another assessment of the plausible mass range for HIP 39017 b and may identify acceptable dynamical solutions that are strongly disfavored.  
We down-sample the \texttt{orvara} results with \textit{informative posteriors}, removing solutions whose companion masses are inconsistent with luminosity evolution analyses.

\subsubsection{Dynamical Modeling Results: Full Posterior Distribution}\label{ss:simplepriors}

We employ a broad informative Gaussian prior for the primary mass, a log-normal prior for mass and semimajor axis, and uninformative priors for other parameters (Table~\ref{orbit_params_hip39017}). The mass we adopt for HIP~39017~A is chosen to be $\rm 1.54\pm0.25\,M_{\odot}$ to appropriately encompass reported literature values and its CMD position as discussed in Section~\ref{sec:obs}. Our PT-MCMC process uses 100 walkers and 30 temperatures across $10^5$ steps.  The convergence diagnostic of our MCMC simulations is based on the auto-correlation time-based criterion, where we evaluate the ratio of the total number of samples to the integrated auto-correlation time (N/$\tau_{int}$), with a threshold value of 50 (ac=50; \citealt{Hogg_2018}). We discard 40$\%$ of the coldest chain as burn-in so that the remaining chain converges adequately for posterior orbital parameter inference.

\begin{figure*}
    \centering
    \begin{overpic}[width=0.7\textwidth]{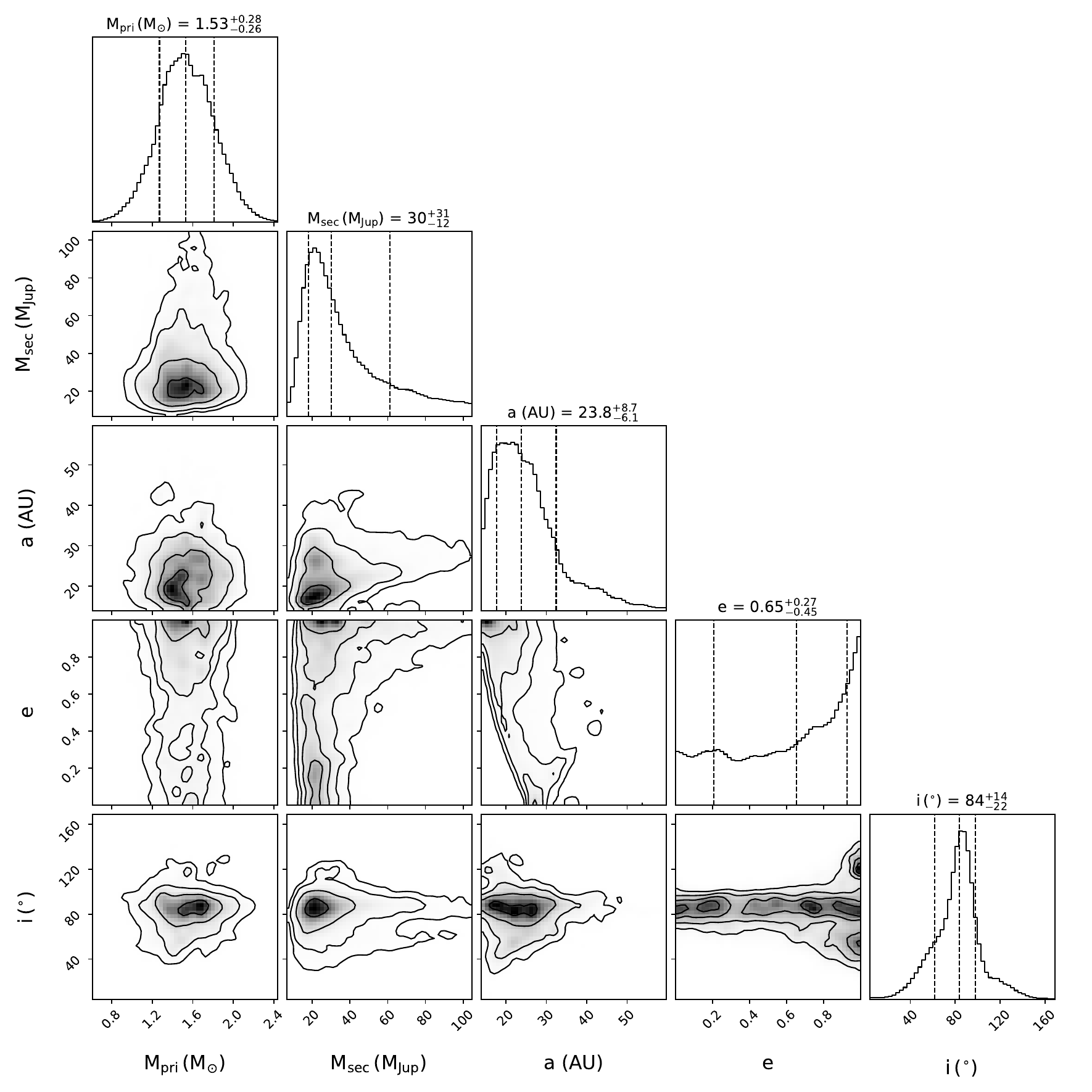}
        \put(65,65){\includegraphics[width=0.35\textwidth]{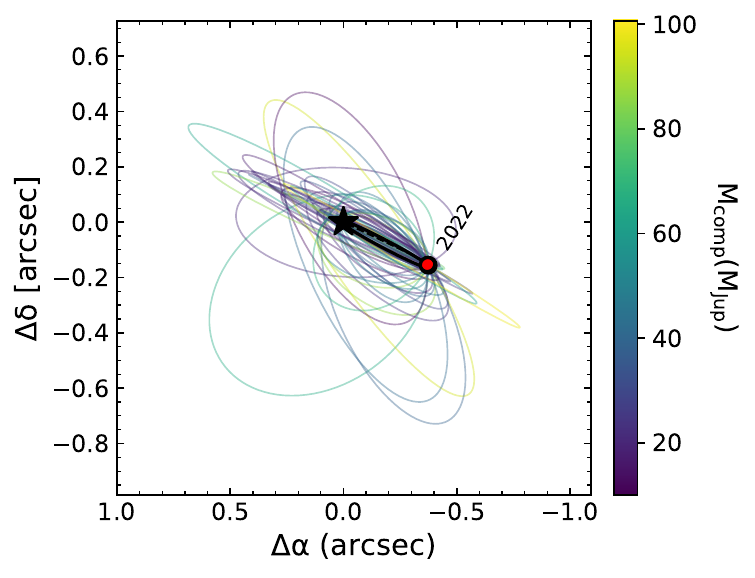}}
    \end{overpic}
    \caption{Posterior distributions of selected parameters from a joint fit with unweighted companion mass posteriors and using the full datasets of relative astrometry from direct imaging and complementary absolute astrometry from the HGCA. The contours show 1, 2, and 3$\sigma$ contour levels at 68$\%$, 95$\%$ and 99.7$\%$ confidence levels.}
    \label{fig:corner_original}
\end{figure*}

The posterior distributions of selected parameters from our simple joint fit to absolute and relative astrometry are shown by a corner plot in Figure~\ref{fig:corner_original}. HIP 39017 b's orbit has a semimajor axis of $\sim$ ${23.8}_{-6.1}^{+8.7}$ au and is likely within $\sim$20$^{\circ}$ of being edge on ($i \sim {84}_{-22}^{+14}$ degrees).  The dynamical mass for HIP ~39017~b retrieved from this raw fit is poorly constrained, with a wide 68\% confidence interval of $M_{\rm comp}$ = ${30}_{-12}^{+31} M_{Jup}$, as is the eccentricity ($e = {0.65}_{-0.45}^{+0.27}$).  The implied mass ratio ($q$ $\sim$ ${1.9\%}_{-0.8}^{+1.9}$).

The companion's mass, mass ratio, and eccentricity from this modeling are highly uncertain because HIP 39017 b was evidently observed during a period when the orbital motion is slow, resulting in the three epochs of relative astrometry being closely spaced and the orbital phase being sparsely sampled. Consequently, the other parameters we derive from the maximum likelihood orbit given in Figure~\ref{fig:corner_original} are also subject to biases.

\begin{deluxetable*}{llll}
\tablewidth{0pt}
\tablecaption{Orbital parameters for HIP~39017~b derived with \texttt{orvara}. \label{orbit_params_hip39017}}
\tablehead{\colhead{Parameter}&
\colhead{Prior} &
\colhead{Posterior} &
\colhead{Weighted Posterior \tablenotemark{a}}} 
    \startdata
\hline
Modeled Parameters\\
\hline
$M_{*}$ ($M_{\odot}$) & N(1.54,0.27)& ${1.53}_{-0.26}^{+0.28}$ & ${1.52}_{-0.26}^{+0.28}$ \\
$M_{p} (M_{Jup}$) & 1/$M_{p}$ & ${30}_{-12}^{+31}$ & ${23.6}_{-7.4}^{+9.1}$  \\
$a$ (AU) & 1/a & ${23.8}_{-6.1}^{+8.7}$ &  ${22.1}_{-4.9}^{+8.0}$ \\
$\sqrt{e} sin(\omega)$ & U(-1,1) & ${0.09}_{-0.64}^{+0.62}$ &  ${0.04}_{-0.51}^{+0.54}$ \\
$\sqrt{e} \cos(\omega)$ & U(-1,1) & ${-0.03}_{-0.61}^{+0.62}$ & ${-0.10}_{-0.60}^{+0.72}$ \\
$i (deg)$ & sini & ${84}_{-22}^{+14}$ & ${84}_{-20}^{+14}$ \\
$\Omega (deg)$ & U(-180,180) & ${223}_{-158}^{+28}$ & ${235}_{-170}^{+14}$  \\
$\lambda_{ref} (deg)$ & U(-180,180) & ${185}_{-118}^{+138}$ & ${195}_{-89}^{+133}$ \\
\hline
Derived Parameters\\
\hline
Parallax (mas)   &  --     & ${15.1782}_{-0.0042}^{+0.0041}$ &  ${15.1782}_{-0.0042}^{+0.0041}$ \\
Period (yrs)    &  --     &      ${93}_{-32}^{+59}$& ${84}_{-26}^{+51}$\\
Argument of periastron (deg)      &  --     &     ${162}_{-102}^{+134}$ &${174}_{-115}^{+131}$ \\
Eccentricity   &  --     &  ${0.65}_{-0.45}^{+0.27}$&     ${0.55}_{-0.39}^{+0.35}$ \\
Semi-major axis (mas)    &  --     & ${361}_{-92}^{+131}$ & ${335}_{-75}^{+122}$ \\
$T_{0}$ (JD)    &  --     &   ${2469973}_{-3921}^{+14463}$ & ${2469682}_{-3741}^{+14648}$  \\
Mass ratio      &  --     &  ${0.0193}_{-0.0082}^{+0.019}$  & ${0.0148}_{-0.0049}^{+0.0065}$\\
    \enddata
     \tablenotetext{a}{Derived using uniform, informative mass posteriors over a range of $\rm M_{p}=$ [0,46] $\rm M_{J}$ anchored by hot-start evolutionary model predictions (see Figure~\ref{fig:evo_tracks}).}
\end{deluxetable*}

\subsubsection{Dynamical Modeling Results: Informed Posteriors} \label{ss:orbitevolpriors}

\begin{figure}
    \centering
\includegraphics[width=0.42\textwidth]{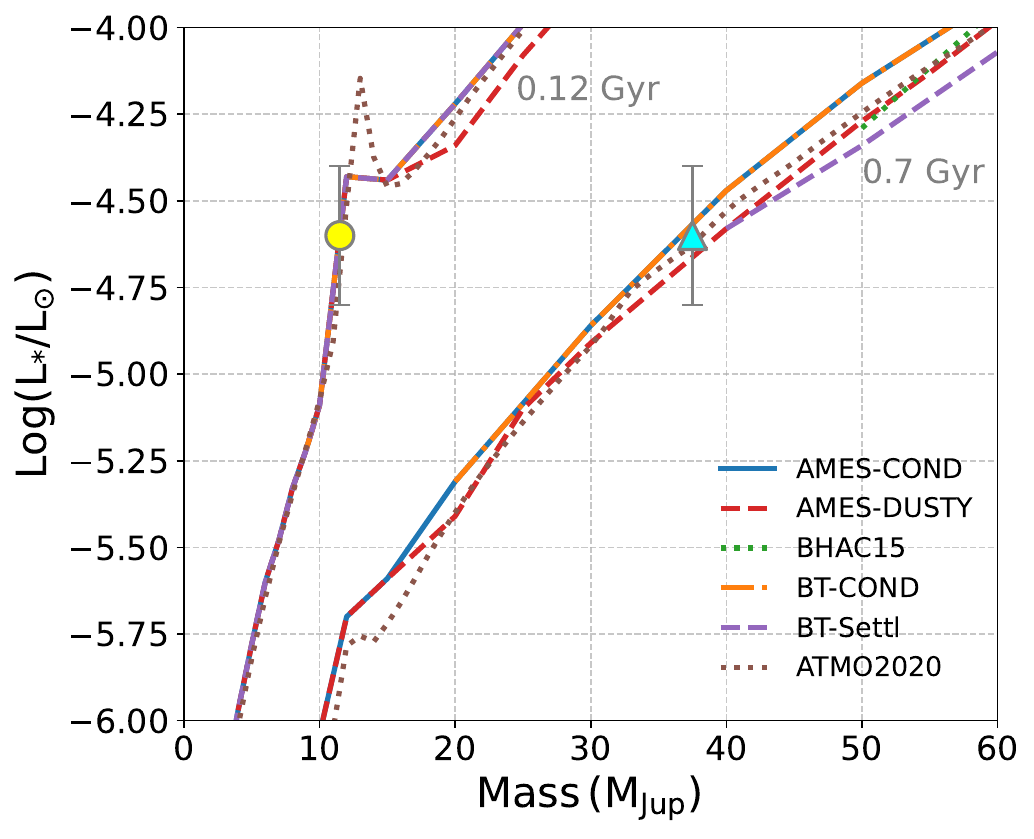}\\
    \caption{ Mass evolution isochrones for 0.12-Gyr and 0.7-Gyr objects given by several different hot-start evolutionary models. The positions of HIP~39017~b are highlighted with yellow and cyan points along the substellar iso-age evolutionary tracks. With an estimated bolometric luminosity of -4.6$\pm$0.2 dex, the predicted masses at ages of 0.12 Gyr and 0.7 Gyr have median values of $13.8^{+1.3}_{-1.6} M_{J}$ and $38.2^{+7.9}_{-5.2} M_{J}$, respectively. The mass uncertainties were estimated by adding the dispersion across the considered evolutionary tracks and the luminosity-related mass uncertainty in quadrature. We adopt 0.7 Gyr/$38.2^{+7.9}_{-5.2} M_{J}$ as the upper age/mass limit. The final conservative upper mass limit from evolutionary models is then 46$M_{J}$. We adopt 0.12 Gyr/$13.8^{+1.3}_{-1.6} M_{J}$ -- the closest grid value to 115 Myr -- as a more realistic characterization for HIP~39017~b. We note that a younger age than 115 Myr may indicate a even lower mass planetary companion as suggested by evolutionary model predictions. 
    }
    \label{fig:evo_tracks}
\end{figure}

To further constrain the orbital mass of the companion, we first investigate the use of information from evolutionary models to downsample the original orbital posteriors derived above.
Given age and bolometric luminosity, substellar evolutionary models predict the mass based on cooling physics. We use bi-linear interpolation to form a grid of mass predictions as a function of age and luminosity. The primary source of uncertainty in evolutionary model-predicted mass estimates is due to the uncertainty in age. We place a conservative upper age limit of $\le 700$ Myr (see Section \ref{ss:systembkgd}) while we obtain the bolometric luminosity from both NIRC2's single-band $L^{\prime}$ photometry and the entire broadband spectrum, by comparing them to observed L and T dwarf spectra. 

We consider two derivations of HIP 39017 b's luminosity. First, we convert the $L^{\prime}$ or W1 band magnitude, $16.12 \pm 0.18$ mag, to an absolute magnitude in W1 using the relation in \citet{Franson2023} and subsequently to a bolometric luminosity following the same approach as \citet{Li2023} using a comparative sample of L and T dwarfs from \citet{Filippazzo_2015}. The bolometric luminosity obtained using this method is  $log(L_{bol}/L_{\odot}) = -4.5\pm 0.1$ dex for HIP~39017~b. 

Second, we consider an atmospheric modeling-derived estimate the bolometric luminosity from the best fit Sonora Diamondback (\citediamondback) and Lacy/Burrows models for HIP 99770b \citep{Currie2023} described above in Section \ref{ss:modspec}. The flux of each best fit model is scaled by the fitted amplitude and integrated over the full extent of wavelengths available for each model ($0.30-250$~$\mu$m for Sonora Diamondback, $0.43-300$~$\mu$m for Lacy/Burrows 99770). The luminosity is then calculated using the flux radius, $R_{F\nu}$, in Table \ref{tbl:spec_fits}. This results in a $log(L_{bol}/L_{\odot}) \sim -4.7$~dex for the best fit Sonora Diamondback model, and $\sim -4.6$~dex for the best fit Lacy/Burrows 99770 model. These estimates are subject to model uncertainties, and could be slightly lower than the true bolometric luminosity due to unmeasured contribution from wavelengths outside of the spectral range of the models. However, they are consistent with the empirically-determined bolometric luminosity of HIP~39017~b determined above.    

Considering both empirical and model-derived estimates, we derive a luminosity of $log(L_{bol}/L_{\odot}) = -4.6 \pm 0.2$ for HIP 39017 b. We compare the measured luminosity to several evolutionary tracks: ATMO2020 \citep{Phillips2020}, AMES-COND \citep{Allard2001}, AMES-DUSTY \citep{Allard2001}, BHAC15 \citep{Baraffe2015}, and BT-Settl \citep{Allard2012, Allard2013}.  Using both the provisional age of 115 Myr and the upper limit age of 700 Myr, we determine the mass inferred from each of these models given HIP 39017 b's luminosity and its associated uncertainty (see Figure \ref{fig:evo_tracks}). The median of these comparisons yields a 
model-implied mass of $13.8^{+1.3}_{-1.6} M_{J}$ at 115 Myr or $38.2_{-5.2}^{+7.9} M_{J}$ at 700 Myr. The corresponding model-inferred parameters are $T_{eff} = 1200_{-120}^{+100} K$, $log(g) = 4.41_{-0.05}^{+0.08}$ dex, and $R = 0.85_{-0.0047}^{+0.53}\, M_{J}$ if assuming an age of 115 Myr. Alternatively, an age of 700 Myr yields parameters of $T_{eff} = 1330_{-150}^{+170} K$, $log(g) = 5.02_{-0.07}^{+0.08}$ dex, and $R = 0.72_{-0.014}^{+0.0092}\, M_{J}$. 

\begin{figure*}
    \centering
\includegraphics[height=0.29\textwidth]{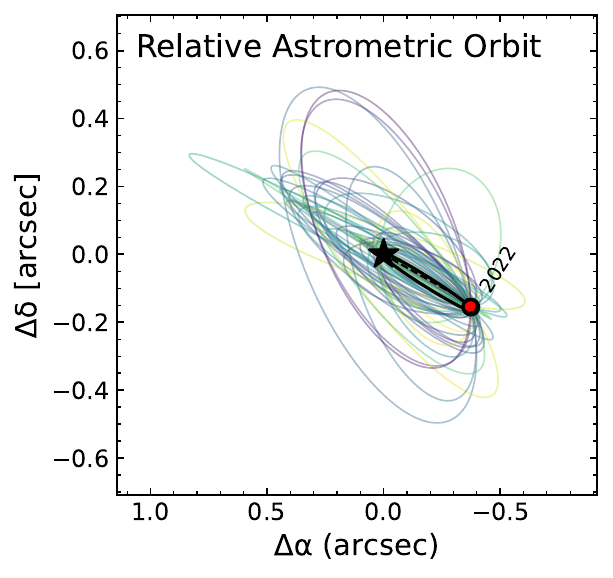}\quad 
\includegraphics[height=0.29\textwidth]{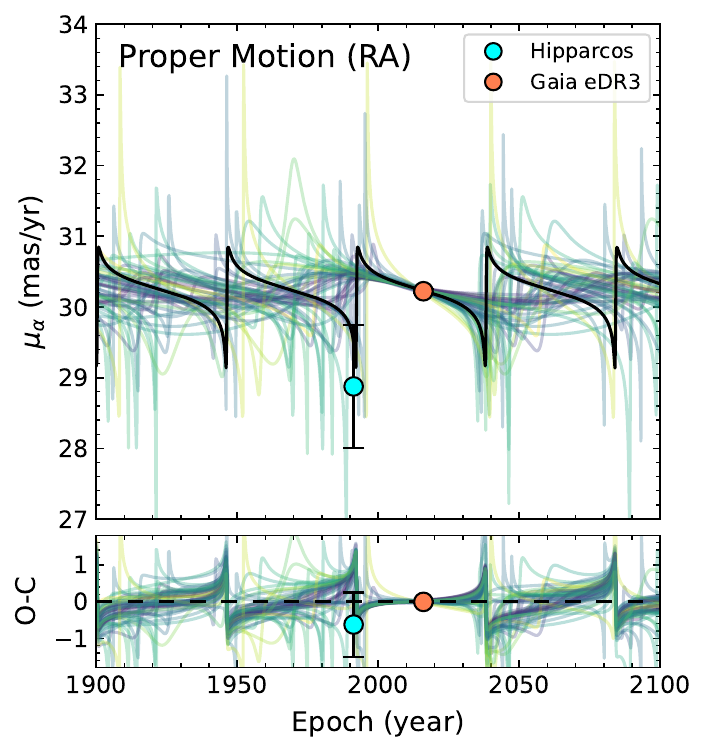}
\includegraphics[height=0.29\textwidth]{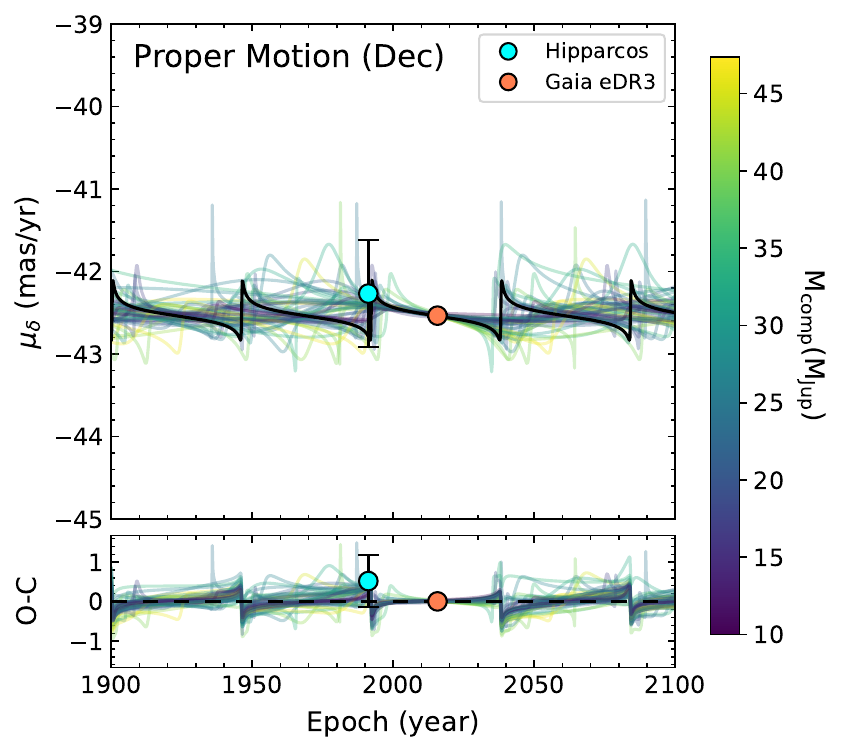}
\caption{Weighted MCMC orbital posteriors using an uniform, informative range of [0,46] $M_{J}$ on the companion mass. This is a result from down-sampling the original fit to the relative astrometry from direct imaging presented in Table~\ref{tbl:det_hip39017} (left panel) and stellar absolute astrometry in R.A. and decl. from the HGCA eDR3 (right two panels). The red data points in the left panel are relative astrometry data driven by direct imaging observations while the two data points in cyan and orange in the right panels denote proper motion measurements from the cross-calibrated Hipparcos and Gaia eDR3 catalogs.\label{fig:orvara_fit_evolpriors}}
\end{figure*}

\begin{figure*}
    \centering
       \vspace{-0.0in}
\includegraphics[width=0.7\textwidth,trim=0mm 0mm 0mm 0mm,clip]{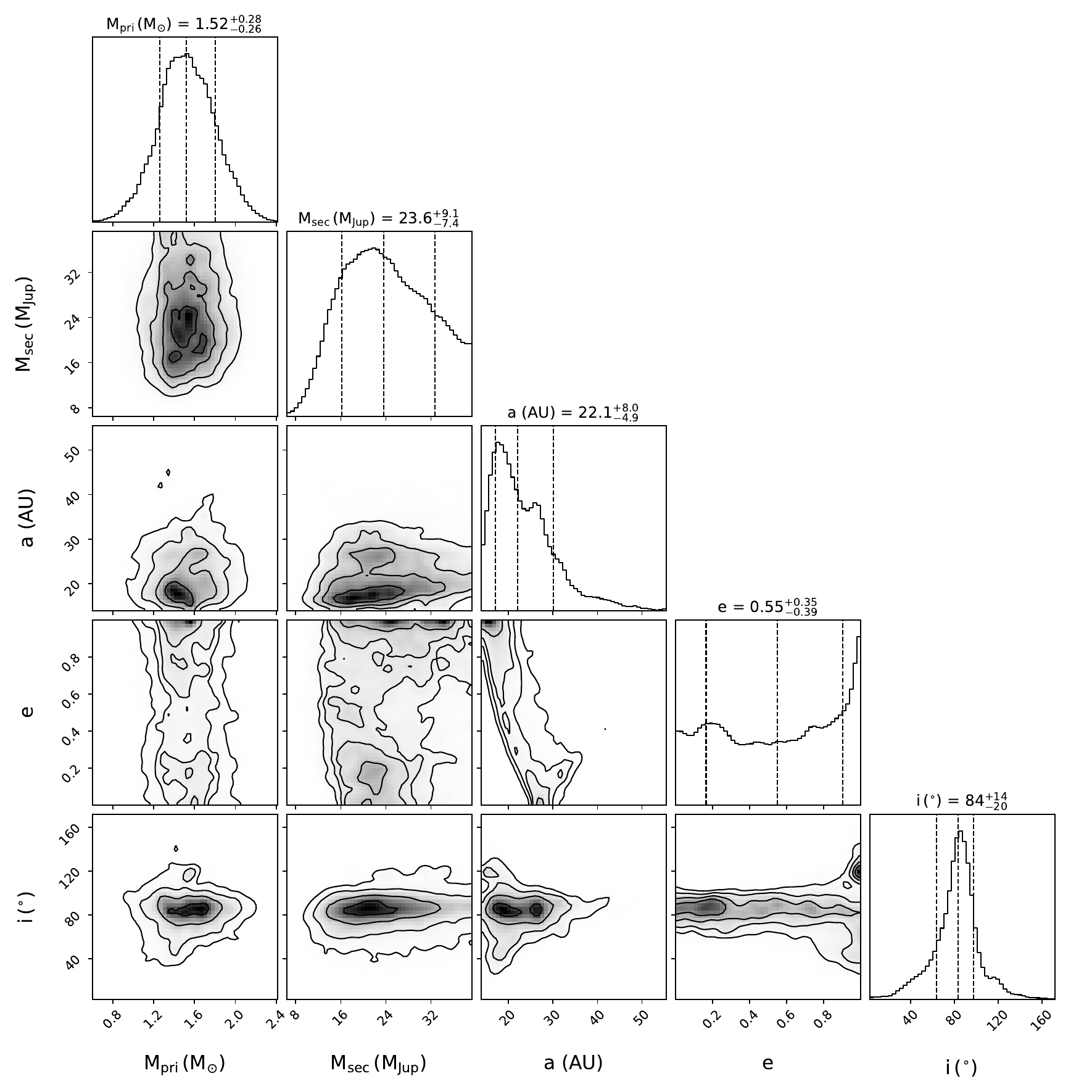}\\
    \vspace{-0.1in}
    \caption{Same as Figure~\ref{fig:corner_original} but employing an uniform, informative cutoff of 0-46 $M_{Jup}$ on the companion mass posteriors by down-sampling the original posteriors in Figure~\ref{fig:corner_original}.}
    \vspace{-0.in}
    \label{fig:corner_evolpriors}
\end{figure*}

The evolutionary-model inferred masses are then used to truncate the posterior orbital parameter distributions for HIP~39017~b by down-sampling orbits within a range between zero and the conservative upper mass limit of 46~$M_{J}$. Figure~\ref{fig:orvara_fit_evolpriors} shows the joint fit to the relative astrometry and absolute astrometry data points after resampling the original fit to the full datasets. The period of the companion is comparable to the 25 yr temporal baseline between Hipparcos and Gaia, which limits our ability to characterize the orbital shape (or eccentricity) and dynamical mass. Posterior distributions of relevant model parameters after down-sampling are shown in Figure~\ref{fig:corner_evolpriors}. A summary of the modeled and derived parameters before and after sampling are given in Table~\ref{orbit_params_hip39017}. 

Compared to the previous fits, the mode of the mass distribution remains unchanged but the long tail of higher masses in conflict with luminosity evolution estimates is truncated.  Thus, while most parameter distributions are consistent with results from the full posterior distribution fit, the dynamical mass is reduced from $30_{-12}^{+31} M_{Jup}$ 
to $23.6_{-7.4}^{+9.1} M_{Jup}$ after down-sampling over the range of masses allowed by evolutionary track predictions. The derived mass ratio drops to $0.0148_{-0.0049}^{+0.0065}$.  In contrast to the full posterior fit, the mass and mass ratio median and 68\% confidence intervals lie below empirically-motivated thresholds -- $<$25 $M_{\rm J}$ and $q$ $<$0.025 -- to identify massive planets and distinguish them from low-mass brown dwarfs \citep{Currie2023,Currie2023b}.

\section{Discussion} \label{sec:disc}

We report the discovery of a substellar companion orbiting the young accelerating $\gamma$ Doradus-type variable star, HIP~39017. Empirical comparisons to HIP 39017 b's photometry and spectrum indicate that it lies at the L/T transition with a best-estimated spectral type of T0.5--T2.   Atmospheric modeling with the Sonara Diamondback grid achieves a good fit and indicates that HIP 39017 b is a cloudy object with a best-estimated temperature, gravity, and radius of $T_{\rm eff}$ = 1300 K, $\log g = 4.0$, and R$ = 0.87 R_{\rm J}$.  Best fit values from the Lacy/Burrows cloudy atmosphere models are similar.

Modeling astrometric data for HIP 39017 b and the primary constrain the companion's semimajor axis to be $\approx$ 23 au and its orbit to be near edge-on, but otherwise yield poor constraints on system properties. The full posterior distribution for HIP 39017 b's dynamical mass and mass ratio have a wide range of $30^{+31}_{-12}$M$_J$ and ${1.9\%}_{-0.8}^{+1.9}$, respectively, implying a brown dwarf interpretation. 
However, given HIP 39017 b's best-estimated age ($\lesssim$ 115 Myr), luminosity evolution models favor a lower mass of $\lesssim$ 13.8 $M_{\rm J}$. Adopting the highly-conservative assumption that HIP 39017's age is simply ``younger than the Hyades," we estimate an upper mass limit of $\approx$ 38 $M_{\rm J}$.   Informed by the latter results -- i.e. removing solutions from the posterior distribution with masses incompatible with luminosity evolution estimates -- the revised posterior distributions for HIP 39017 b yield a lower mass of $23.6_{-7.4}^{+9.1} M_{Jup}$ and a lower mass ratio of $0.0148_{-0.0049}^{+0.0065}$. These combined values place this companion in a region of the parameter space populated by massive planets detected by RV and direct imaging, favoring a planetary interpretation as described in \citet{Currie2023,Currie2023b}.

Current CHARIS and NIRC2 data for HIP 39017 b suggest that our images capture a part of its orbit where it undergoes little astrometric motion.  Continued astrometric monitoring of HIP 39017 b over a long time baseline will substantially improve constraints on its orbit and, potentially, its dynamical mass. The Gaia Data Release 4 will yield more precise astrometric data for the star, which will further improve mass estimates and thus clarify its nature (i.e. whether it shares more of its properties with low-mass brown dwarfs or planets). 

A more precise age for the system is needed for comparison to evolutionary models. Our estimate of $\lesssim 115$~Myr implies a lower mass of $\lesssim 13.8 M_J$ from comparison with several hot-start evolutionary models (Figure \ref{fig:evo_tracks}), placing HIP~39017b closer to the deuterium burning limit. Its host star's multi-frequency, $\gamma$ Dor-type variability opens the door to deriving independent age constraints for the system through asteroseismology. Further observations with higher spectral resolution would allow the derivation of stronger constraints on its spectral properties (e.g. $T_{eff}$, $\log g$, metallicity, C/O ratio) via atmospheric retrieval.

The discovery of HIP~39017b reinforces the improved efficiency of targeted direct imaging campaigns informed by long-baseline, precision stellar astrometry instead of so-called \textit{blind} or unbiased surveys. Between the system's youth and potential for independent dynamical mass constraints, HIP~39017b provides a new testing ground for substellar evolutionary and atmospheric models. 
HIP 39017 b joins recently-discovered planetary companions HIP 99770 b and AF Lep b and brown dwarfs  HIP 21152 B and HD 33632 Ab as companions with spectra that can constrain their atmospheres and direct dynamical mass measurements. With improved age constraints for all companions, new targeted surveys will establish a sample of companions that will help map how the atmospheres of planets and brown dwarfs of a given mass evolve with time.

\begin{acknowledgments}
\indent We thank Caroline Morley for providing the Sonora Diamondback model grids, and we thank the referee for their thoughtful comments. 

The authors wish to recognize and acknowledge the very significant cultural role and reverence that the summit of Maunakea has always had within the Native Hawaiian community. We are honored and grateful for the opportunity to conduct observations from Maunakea. 
This research is based in part on data collected at the Subaru Telescope, which is operated by the National Astronomical Observatory of Japan. The development of SCExAO is supported by the Japan Society for the Promotion of Science (Grant-in-Aid for Research \#23340051, \#26220704, \#23103002, \#19H00703, \#19H00695 and \#21H04998), the Subaru Telescope, the National Astronomical Observatory of Japan, the Astrobiology Center of the National Institutes of Natural Sciences, Japan, the Mt Cuba Foundation and the Heising-Simons Foundation. CHARIS was built at Princeton University under a Grant-in-Aid for Scientific Research on Innovative Areas from MEXT of the Japanese government (\#23103002). 
\end{acknowledgments}\begin{acknowledgments}
Some of the data presented herein were obtained at Keck Observatory, which is a private 501(c)3 non-profit organization operated as a scientific partnership among the California Institute of Technology, the University of California, and the National Aeronautics and Space Administration. The Observatory was made possible by the generous financial support of the W. M. Keck Foundation. 
This research has benefitted from the SpeX Prism Spectral Libraries, maintained by Adam Burgasser at \url{https://cass.ucsd.edu/~ajb/browndwarfs/spexprism}. This research has made use of the SIMBAD  \citep{Wenger2000} and VizieR \citep{Ochsenbein2000} databases, operated at CDS, Strasbourg, France. 
\end{acknowledgments}

\facilities{
    Subaru (SCExAO, CHARIS),
    Keck:II (NGS-AO, NIRC2)
}

\software{
    \texttt{ADEPTS} \citep{Tobin2020, Tobin2022},
    \texttt{Astropy} \citep{astropy1, astropy2},
    \texttt{Astropy Specutils FluxConservingResampler} \citep{specutils,FluxConservingResampler},
    CHARIS Data Processing Pipeline (\texttt{DPP}; \citealt{Currie2020b}),
    CHARIS Data Reduction Pipeline (\texttt{DRP}; \citealt{Brandt2017}),
    \texttt{emcee} \citep{Foreman-Mackey_2013},
    \texttt{lacosmic} \citep{vanDokkum2001},
    \texttt{NumPy} \citep{numpy},
    \texttt{orvara} \citep{Brandtorvara},
    \texttt{SciPy} \citep{scipy},
    SpeX Prism Library Analysis Toolkit (\texttt{SPLAT}; \citealt{splat}),
    \texttt{Matplotlib} \citep{matplotlib,matplotlib3.4.3},
    \texttt{tol-colors} \citep{tol-colors-code,tol-colors-technote},
    Vortex Image Processing library (\texttt{VIP}; \citealt{Gomez2017}
}

\bibliography{references}{}
\bibliographystyle{aasjournal}

\appendix 
\section{Characterization of a Wide-Separation Background Star Detected Around HIP 39017}
\label{apx:bkgd-src}
\begin{figure}
             \includegraphics[width=0.325\textwidth,trim=0mm 0mm 0mm 0mm,clip]{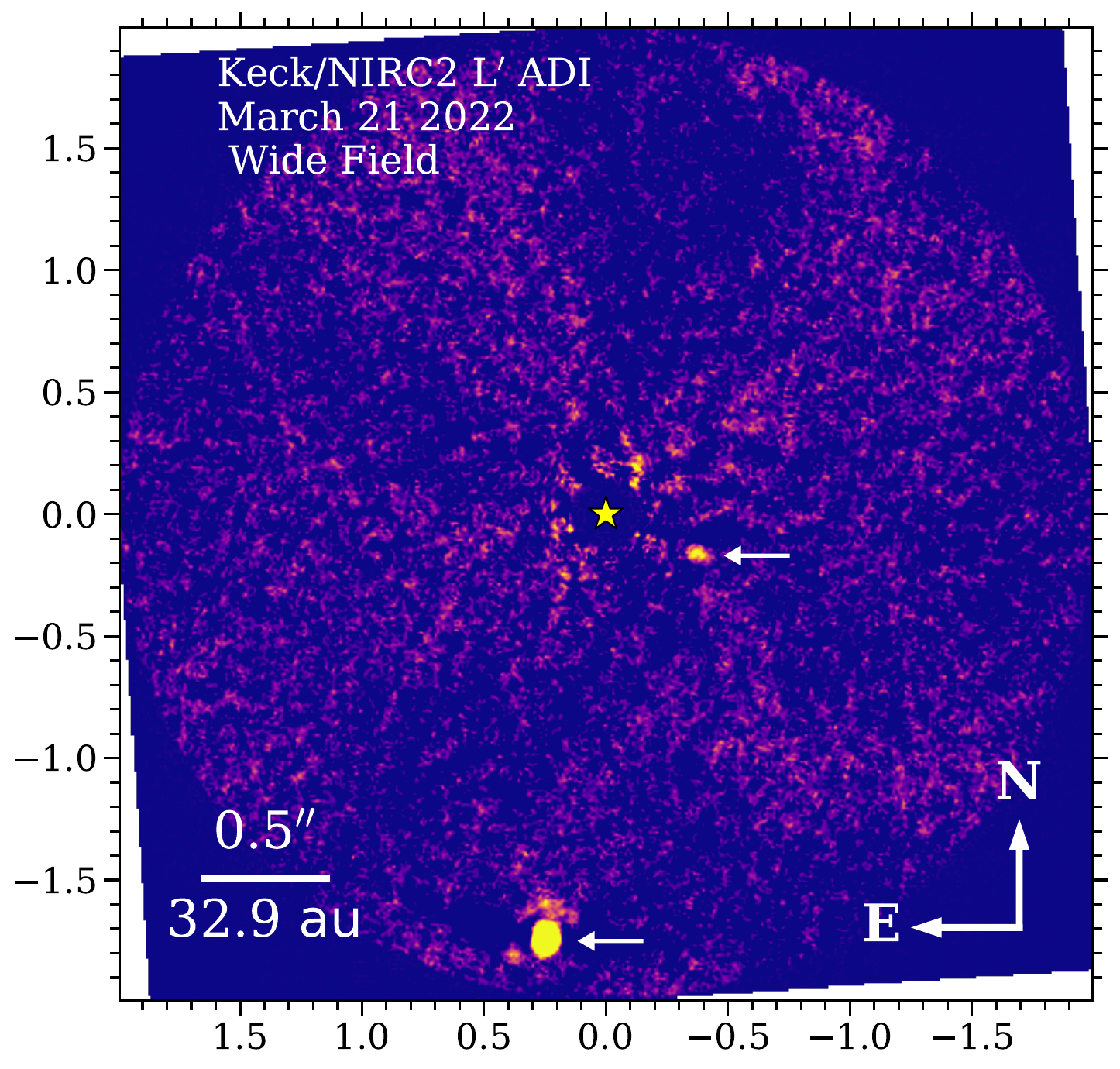}
      \includegraphics[width=0.325\textwidth,trim=0mm 0mm 0mm 0mm,clip]{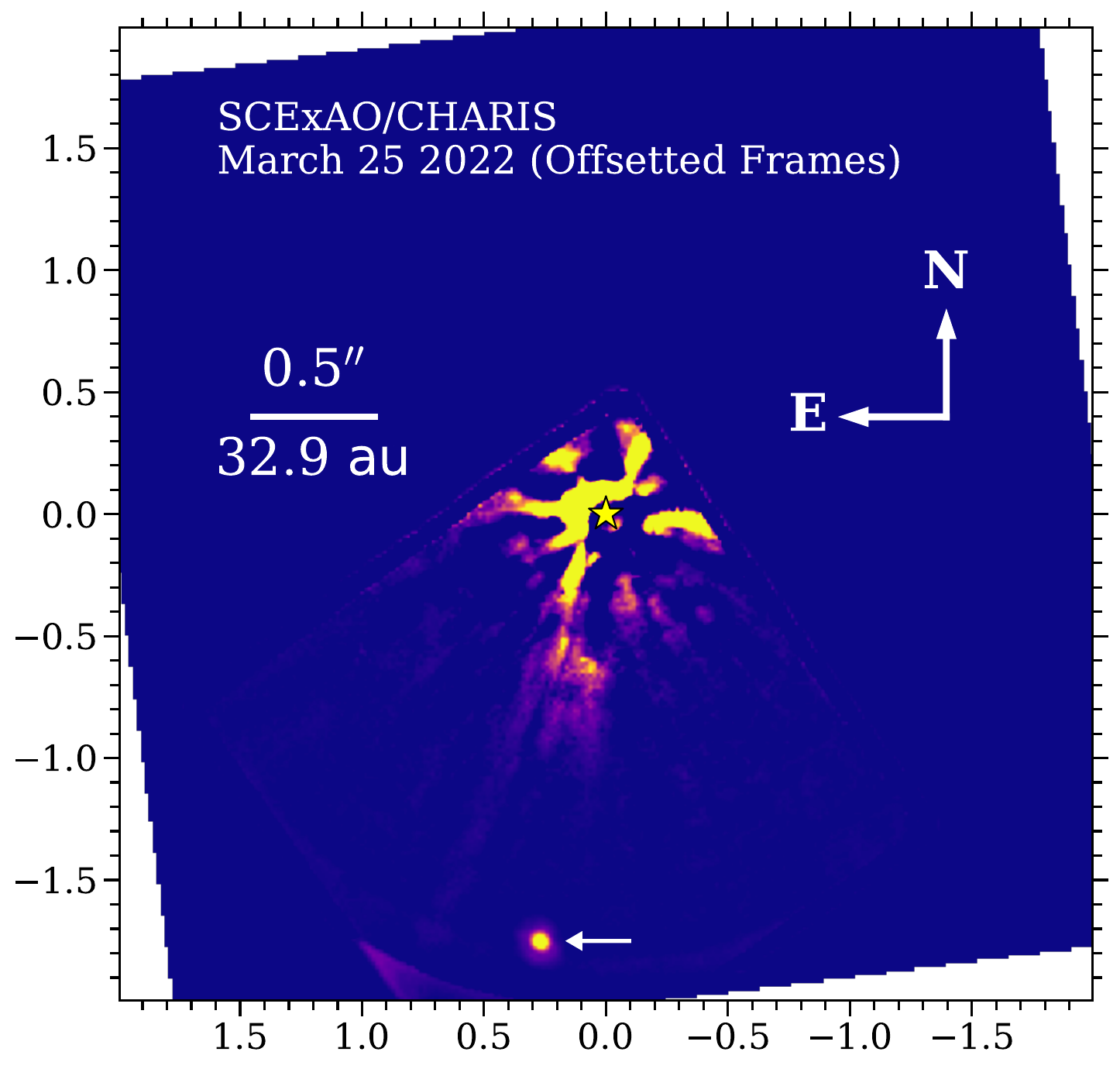}
      \includegraphics[width=0.325\textwidth,trim=0mm 0mm 0mm 0mm,clip]{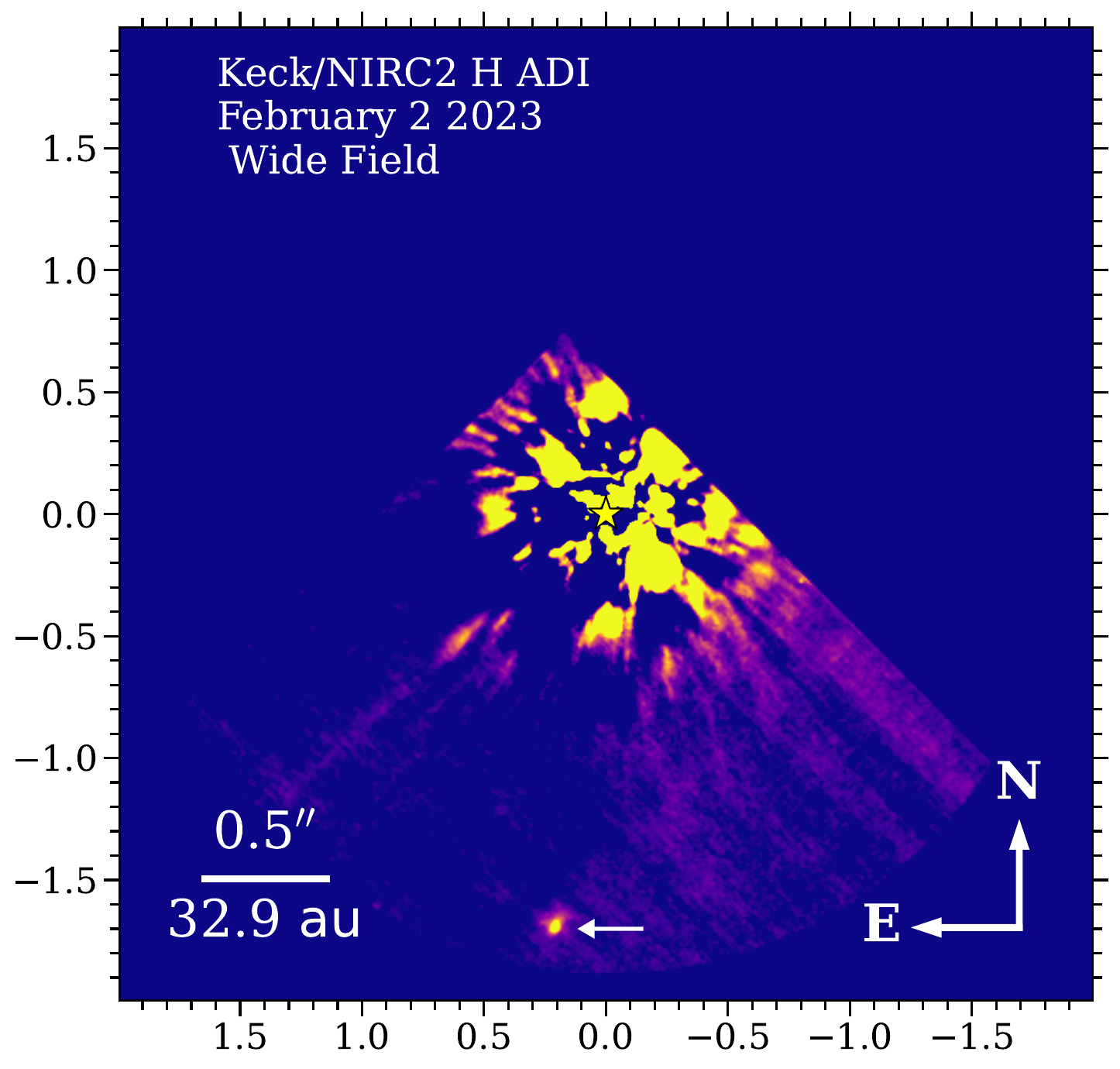}
    \vspace{-0.1in}
    \caption{Detections of a wider-separation point source around HIP 39017 with Keck/NIRC2 (left and right panels) and SCExAO/CHARIS (middle panel)}
    \vspace{-0.in}
    \label{fig: bckgdstarimage}
\end{figure}

The 2022 NIRC2 data identify  a second point source in the northwest at a wider separation ($\rho$ $\sim$ 1\farcs{}75), which is recovered in the March 2022 CHARIS epoch and 2023 NIRC2 epoch (Figure \ref{fig: bckgdstarimage}).   If a bound companion at a projected physical separation of 115 au, this object would contribute to the HIP 39017 primary's astrometric acceleration and thus substantially the orbital solutions for HIP 39017 b.      However, our analysis below shows that this object is an unrelated background object whose colors are consistent with an unreddened K star.

Table \ref{tbl:det_hip39017bckgd} lists astrometric measurements for this object.  While basic processing for each data set listed follows methods described in Section 2.2, we did not apply PSF subtraction techniques to detect the wider-separation object in the March 2022 CHARIS and March 2023 NIRC2 data.  For these data sets, we extract an object spectrum and/or measure its astrometry directly.  The detection signifances are sufficiently high (SNR $\sim$ 47-222) that systematics in the plate scale and north position angle calibration for CHARIS and NIRC2 dominate our astrometric error budget.  

\begin{figure}
    \centering
       \vspace{-0.0in}
      \includegraphics[width=0.5\textwidth,trim=0mm 0mm 0mm 0mm,clip]{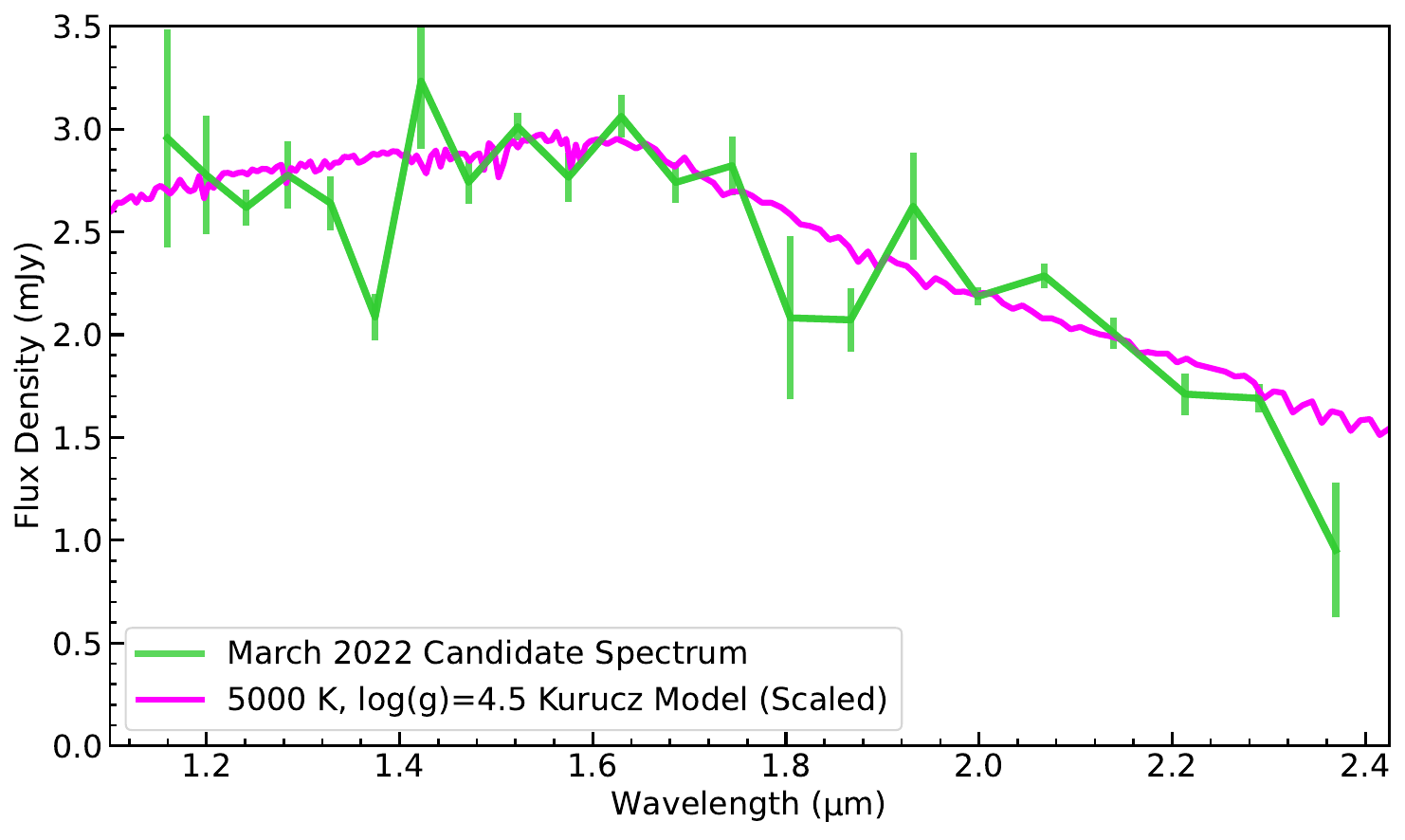}
        \includegraphics[width=0.44\textwidth,trim=0mm 0mm 0mm 0mm,clip]{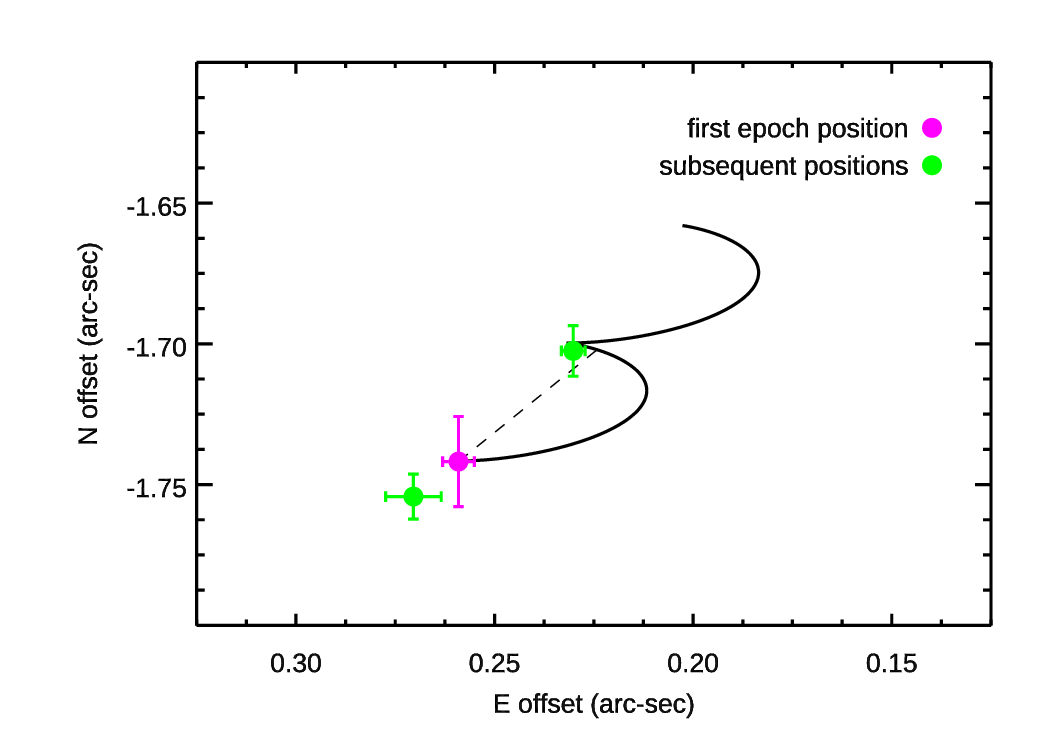}
    \vspace{-0.1in}
    \caption{\textit{Left:} CHARIS spectrum of the wider-separation point source shown to be a background star (green) with a Kurucz model atmosphere corresponding to a K2-K3 star overplotted (magenta). \textit{Right:} Proper motion analysis showing that this point source's astrometry is consistent with the motion expected for a background star.   The dashed line connects the object's position in the first detected epoch to its predicted position in the 2023 Keck/NIRC2 data if it is a background star, a prediction that matches the matches the observations within errors (upper-right green dot).}
    \vspace{-0.in}
    \label{fig:hip39017spectrumbckgd}
\end{figure}

The wider-separation object (Figure \ref{fig:hip39017spectrumbckgd}) shows a spectrum that is smooth and featureless except for channels with substantial telluric contamination (1.3--1.4 $\mu m$, 1.8--1.9 $\mu m$).   We measure its broadband photometry to be $J$ = 14.38 $\pm$ 0.08, $H$ = 13.89 $\pm$ 0.04, and $K_{\rm s}$ = 13.81 $\pm$ 0.04.   At $L^{\prime}$, we measure a brightness of 13.83 $\pm$ 0.04 magnitudes.   The resulting $J$-$K_{\rm s}$ color is 0.57 $\pm$ 0.09, roughly consistent with a background K2--K3 star \citep{PecautMamajek2013} at a distance of $\approx$850 pc.   Aside from channels with significant telluric contamination, the 5000 K, $\log g = 4.5$ Kurucz atmosphere model overplotted on Figure \ref{fig:hip39017spectrumbckgd} matches this object's spectrum.   Similarly, a proper motion analysis shows that this object's motion relative to HIP 39017 A is consistent with a background star


\begin{deluxetable}{llllllllll}
     \tablewidth{0pt}
    \tablecaption{HIP 39017 Background Star Detections\label{tbl:det_hip39017bckgd}}
    \tablehead{
    \colhead{UT Date} &
    \colhead{Instrument\tablenotemark{a}} &
    \colhead{Bandpass} &
    \colhead{S/N} & 
    \colhead{$\rho$} &
    \colhead{PA} 
    \\ 
    \colhead{} &  
    \colhead{} &  
    \colhead{} & 
     \colhead{} &
     \colhead{(mas)} &
     \colhead{(deg)} 
    }
    \startdata
    2022-03-21 & Keck/PyWFS+NIRC2 &  $L^{\prime}$ & 81 & 1761 $\pm$ 3.0 & 171.540 $\pm$ 0.520 \\
    2022-03-25& SCExAO/CHARIS & Broadband/$JHK$ & 222 & 1774 $\pm$ 7.0 & 171.232 $\pm$ 0.270 \\
      2023-02-04& Keck/NIRC2 & $H$ & 47 & 1718 $\pm$ 3.0& 172.300 $\pm$ 0.300\\
    \enddata
      \tablenotetext{a}{The wavelength range for CHARIS is 1.16--2.37 $\mu$m, while the $L^{\rm{\prime}}$-filter's central wavelength for NIRC2 is 3.78 $\mu$m.}
    \tablecomments{
    S/N represents the object's signal-to-noise ratio, $\rho$ represents the companion.}
\end{deluxetable}

\section{Spectral Analysis: Comparison to Cloudless Sonora Bobcat Grid}
\label{apx:dec-bobcat}

\begin{figure}
    \centering
    \includegraphics[height=2.58in]{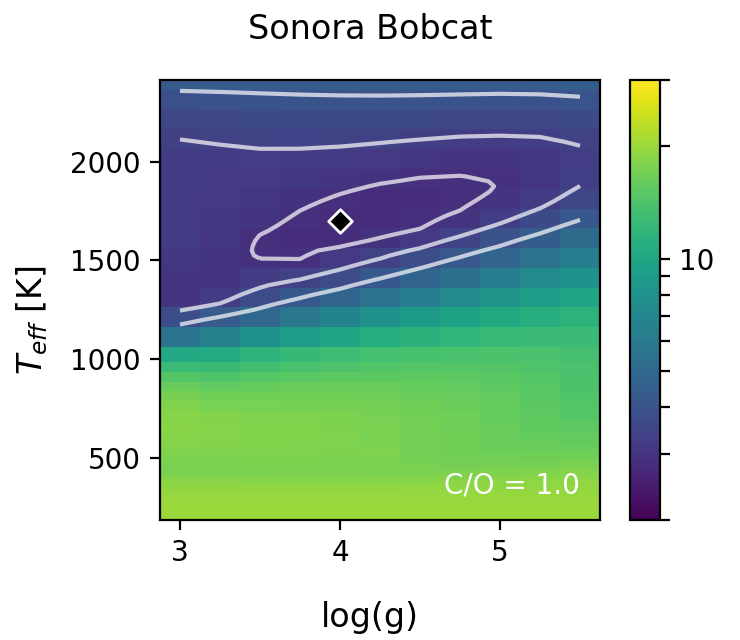}
    \includegraphics[height=2.5in]{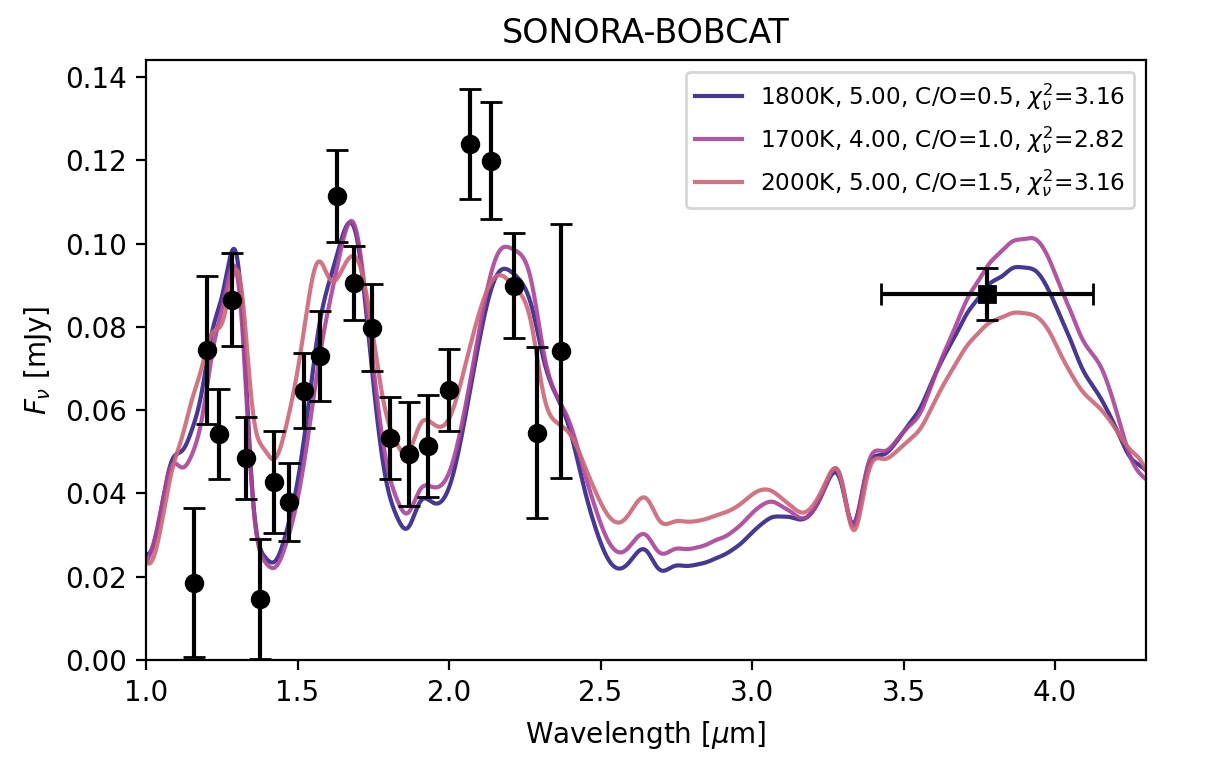}
    \caption{The Sonora Bobcat \citep{Marley2021} fit results to the HIP~39017b spectrum using the December 2022 CHARIS dataset and the Keck/NIRC2 $L^{\prime}$ photometric point. (\textit{Left:}) Map of the $\chi_{\nu}^2$ as a function of model $\log g$ and $T_{eff}$ for Sonora Bobcat models with solar metallicity and C/O. The best fit Sonora Bobcat model is marked with the black diamond, with 1, 2, and 3$\sigma$ contours around the minimum. (\textit{Right:}) The best fit model spectrum (smoothed to $R=50$ at 2.65$\mu$m) for each C/O value provided in the Sonora Bobcat model grid overplotted with the HIP~39017b observations. The legend denotes the $T_{eff}$, $\log g$, C/O, and $\chi_{\nu}^2$ for each model spectrum shown. Note: Sonora Bobcat spectra with C/O$= \{0.5, 1.5\}$ are only available with $\log g=5.0$. }
    \label{fig:bobcat}
\end{figure}

As discussed in Section \ref{ss:modspec}, the results of fitting the December 2022 SCExAO/CHARIS data set and Keck/NIRC2 L$\prime$ photometry to the Sonora Bobcat model grid are shown in Figure \ref{fig:bobcat}. 

\section{Spectral Analysis of the February 2022 SCExAO/CHARIS Spectrum}
\label{apx:feb-spec}

\begin{figure}
    \centering
    \includegraphics[width=0.47\textwidth]{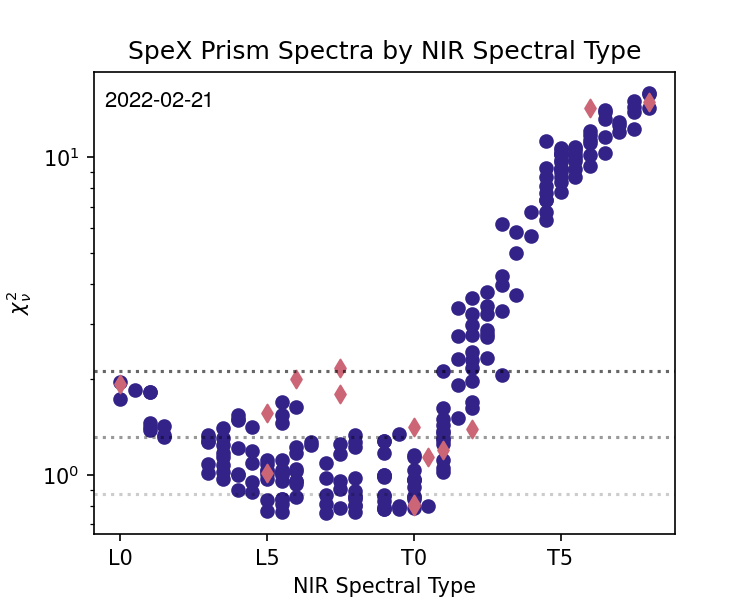}
    \caption{Same as Figure \ref{fig:spex_chi2} but using the $\chi_{\nu}^2$ derived from fitting the empirical spectra to the February 2022 CHARIS spectrum rather than the December 2022 data.}
    \label{fig:spex_chi2_feb2022}
\end{figure}

For comparison, the fits to the SpeX Prism L/T dwarf library and the Sonora Diamondback model grid discussed in Sections \ref{ss:empspec} and \ref{ss:modspec}, respectively, were also performed using the February 2022 SCExAO/CHARIS spectrum. As seen in Figure \ref{fig:hip39017spectrum}, the shape of the $\sim 2.1 \mu$m peak in the February 2022 spectrum more closely resembles that of, for example, the empirical SpeX spectra in Figure \ref{fig:spex_spectra} than the December 2022 spectrum did. However, the weaker J-band peak in the December spectrum is obscured in the February spectrum. In general, we find that the parameters derived using the February spectrum are more poorly constrained and the data are somewhat overfit. Therefore, we defer to the parameters derived from the December spectrum. However, for completeness, we briefly discuss the analysis of the February spectrum here.

Consistent with the increased spectral noise of this data set (Section \ref{sec:spec}), the NIR spectral type implied by the SpeX Prism library spectra is more poorly constrained for the February 2022 data than for the December 2022 data. As can be seen in Figure \ref{fig:spex_chi2_feb2022}, it is consistent with spectral types from $\sim$L5-T1 within $1\sigma$, or from $\sim$L2-T2 within $2\sigma$. Further, the higher uncertainty on this spectrum results in the observations being overfit by the single parameter fit to the library spectra, with many having $\chi_{\nu}^2 < 1$. 

\begin{figure}
    \centering
    \includegraphics[width=0.8\textwidth]{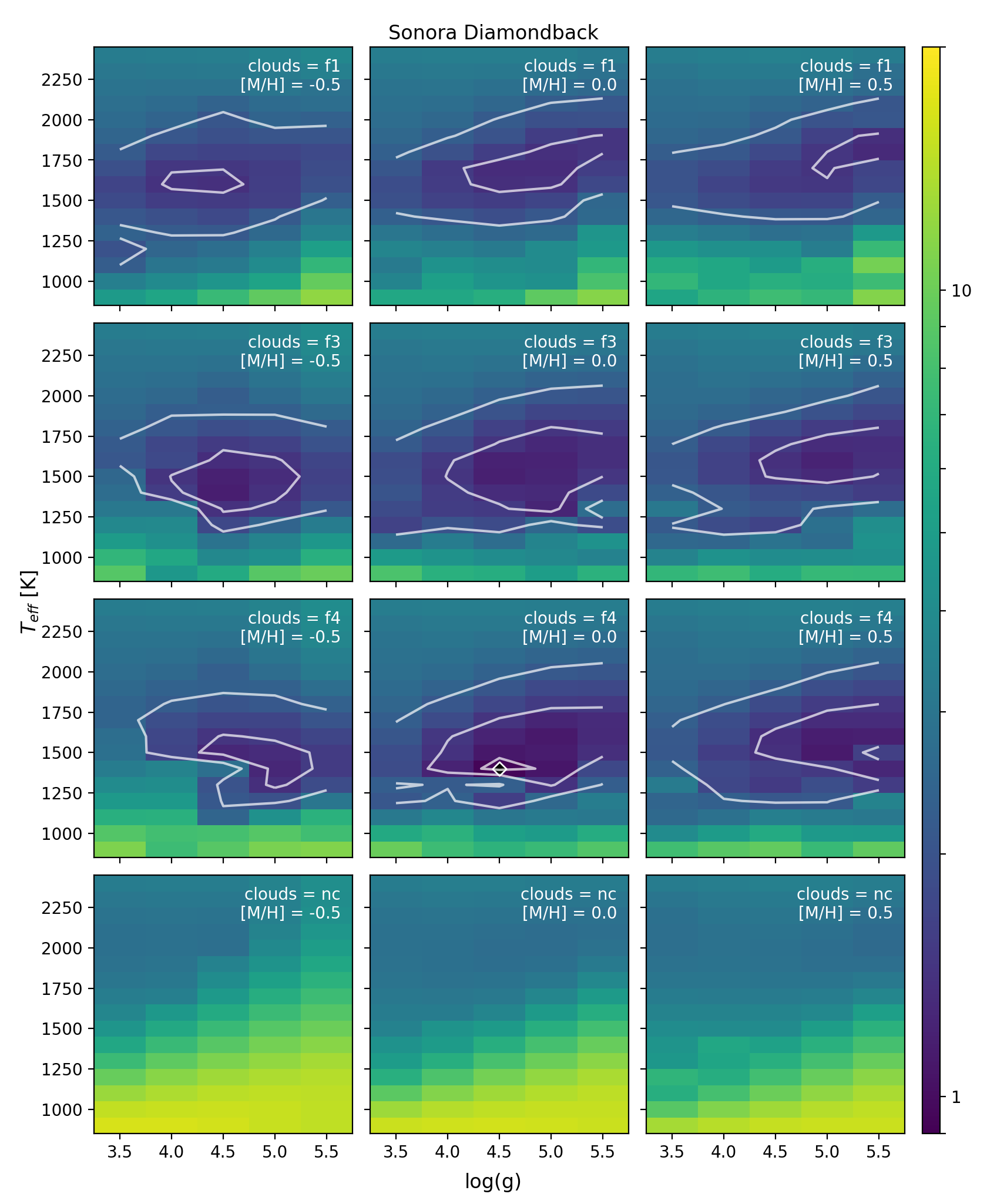}
    \caption{Similar to Figure \ref{fig:diamondback_chi2map}, but using the February 2022 CHARIS spectrum rather than the December 2022 data to fit the Sonora Diamondback model grid.}
    \label{fig:diamondback_chi2map_feb2022}
\end{figure}

The February CHARIS spectrum, in combination with the Keck/NIRC2 $L^{\prime}$ photometry, is also slightly overfit by the Sonora Diamondback grid, but to a much lesser extent, with the minimum $\chi_{\nu}^2 = 0.94$ (Figure \ref{fig:diamondback_chi2map_feb2022}). The Diamondback model that best fits the February 2022 spectrum and $L^{\prime}$ photometry is $T_{eff} = 1400$~K, $\log g = 4.5$, [M/H]$=0.0$, with $f_{sed}=4$ clouds. This is adjacent to the best fit Diamondback model using the December spectrum, but slightly higher temperature and $\log g$.

\end{document}